\pdfoutput=1

\documentclass[final
]{frontiersSCNS} 
\usepackage[T1]{fontenc} 
\input{glyphtounicode} \pdfgentounicode=1
\usepackage[utf8]{inputenx}

\newcommand*{\pdftitle}{Perfect spike detection via time reversal}

\usepackage[onehalfspacing]{setspace}
\usepackage{textcomp}
\usepackage{amsfonts}
\usepackage{upgreek}
\usepackage{bm}
\usepackage{mathtools}
\usepackage{verbatim}
\usepackage{algorithm}
\usepackage[noend]{algpseudocode}
\usepackage{framed}

\usepackage[shortlabels,inline]{enumitem}
\SetEnumitemKey{para}{itemindent=\parindent,leftmargin=0pt,listparindent=\parindent,parsep=0pt,itemsep=\topsep}
\setlist[enumerate,2]{label=\alph*.}
\setlist[enumerate]{leftmargin=\parindent,labelsep=0.5\parindent,topsep=0pt}
\setlist[itemize]{leftmargin=\parindent,labelsep=0.5\parindent,topsep=0pt}

\usepackage{latexsym}
\usepackage{pdfpages}
\usepackage{graphicx}
\usepackage{hyperref}
\usepackage[depth=4]{bookmark}
\providecommand{\href}[2]{#2}

\usepackage{xcolor}
\definecolor{shadecolor}{gray}{0.9}

\definecolor{darkblue}{rgb}{0.0,0.0,0.64}

\definecolor{choose_color}{HTML}{B0CFEC}
\definecolor{connect_color}{HTML}{EDAF9B}
\definecolor{rest_color}{HTML}{EBD7A0}

\usepackage[british]{babel}
\usepackage[autostyle=true,autopunct=false,english=american]{csquotes}
\DeclareQuoteAlias[american]{english}{british}
\setquotestyle{american}

\newcommand*{\chap}{ch.} 
\newcommand*{\sect}{\S} 

\newcommand*{\defd}{\coloneqq}
\newcommand*{\e}{\mathrm{e}}
\newcommand*{\di}{\mathrm{d}}
\newcommand*{\st}{\mathpunct{|}}  
\newcommand*{\Liff}{\mathrel{\ \Leftrightarrow\ }}

\let\oldsum\sum
\renewcommand*{\sum}{\mathop{\textstyle\oldsum}}
\renewcommand*{\prod}{\mathop{\textstyle\oldprod}}
\DeclarePairedDelimiter\set{\{}{\}}

\newcommand*{\tm}{\tau} 
\newcommand*{\ts}{\tau_{\text{s}}} 
\newcommand*{\Is}{I} 
\newcommand*{\Ix}{I_{\text{e}}} 
\newcommand*{\Cm}{C} 
\newcommand*{\ir}{I_{\yvt}} 
\newcommand*{\etmm}{\e^{-\frac{t}{\tm}}} %
\newcommand*{\etsm}{\e^{-\frac{t}{\ts}}} %
\newcommand*{\etmp}{\e^{\frac{t}{\tm}}} %
\newcommand*{\etsp}{\e^{\frac{t}{\ts}}} %
\newcommand*{\ehsm}{\e^{-\frac{\yh}{\ts}}} %
\newcommand*{\ehmp}{\e^{\frac{\yh}{\tm}}} %
\newcommand*{\ehsp}{\e^{\frac{\yh}{\ts}}} %
\newcommand*{\Vres}{V_{\text{r}}} 
\newcommand*{\tauref}{\tau_{\text{r}}} 

\renewcommand*{\textohm}{\ensuremath{\Omega}} 
\newcommand*{\umV}{\;\textrm{mV}} 
\newcommand*{\ums}{\;\textrm{ms}} 
\newcommand*{\upF}{\;\textrm{pF}} 

\newcommand*{\sda}{lossless method}
\newcommand*{\psc}{post-synaptic current}
\newcommand*{\spike}{spike} 
\newcommand*{\nospike}{no-spike}  
\newcommand*{\env}{envelope} 
\newcommand*{\Sp}{\ensuremath{\textit{S}}}
\newcommand*{\NSp}{\ensuremath{\textit{NS}}}

\DeclareMathOperator{\erf}{erf}
\DeclareMathOperator{\Ord}{O}
\newcommand*{\yA}{\mathte{A}}
\newcommand*{\idm}{\mathte{I}}
\newcommand*{\ym}{\bm{q}}
\newcommand*{\ys}{\bm{s}}
\newcommand*{\ysd}{\Dot{\ys}}
\newcommand*{\yso}{\ys_0}
\newcommand*{\RR}{\mathbf{R}}
\newcommand*{\yV}{V} 
\newcommand*{\yI}{\bm{I}}
\newcommand*{\yM}{S}
\newcommand*{\yvt}{\theta}

\newcommand*{\yh}{h} 

\newcommand*{\yto}{t}
\newcommand*{\yal}{\alpha}

\newcommand*{\yarr}{c}
\newcommand*{\yar}{\bm{\yarr}\T}
\newcommand*{\yac}{\bm{d}}
\newcommand*{\yB}{\mathte{B}}
\newcommand*{\ymu}{\beta}
\newcommand*{\yvar}{\sigma^2}
\newcommand*{\yj}{J}
\newcommand*{\yrc}{\bm{r}}
\newcommand*{\yuu}{x}
\newcommand*{\yu}{\bm{\yuu}}
\newcommand*{\yvv}{y}
\newcommand*{\yv}{\bm{\yvv}}
\newcommand*{\yze}{\bm{0}}
\newcommand*{\yN}{N}
\newcommand*{\ybt}{E}
\newcommand*{\ybo}{\bm{\varGamma}}
\newcommand*{\yS}{\chi}
\newcommand*{\yC}{b}
\newcommand*{\yD}{D_{\yS}}
\newcommand*{\Tg}{\mathrm{T}}
\newcommand*{\ylam}{\lambda}
\newcommand*{\ygM}{\mathte{M}}
\newcommand*{\ygv}{\bm{m}}
\newcommand*{\ygr}{\bm{k}\T}
\newcommand*{\ygs}{\kappa}
\newcommand*{\ygrb}{{\bm{k}'}\T}
\newcommand*{\ygsb}{\kappa'}

\newcommand*{\ySi}{X}
\newcommand*{\ydt}{\Delta t} 
\newcommand*{\ytte}{t_{\yvt}} 
\newcommand*{\ytm}{t_{\text{max}}} 

\newcommand*{\eqn}{eq.}
\newcommand*{\eqns}{eqs}
\newcommand*{\alg}{Algorithm}
\renewcommand*{\le}{\leqslant}
\renewcommand*{\ge}{\geqslant}
\newcommand*{\T}{^\intercal}
\DeclarePairedDelimiter\clcl{[}{]}

\DeclarePairedDelimiter\opcl{]}{]}

\DeclarePairedDelimiter\abs{\lvert}{\rvert}
\newcommand*{\mathte}[1]{\ensuremath{\textbf{\textit{\textsf{#1}}}}}
\newcommand*{\ie}{i.e.}
\newcommand*{\eg}{e.g.}
\newcommand*{\etc}{etc.}

\newcommand*{\vs}{vs}

\newcommand{\mynote}[2]%
{\mbox{}\newline\colorbox{#1}{\parbox{0.97\textwidth}{\small #1: #2}}\newline}

\definecolor{notecolour}{RGB}{136,34,85}

\newcommand{\ednote}[2]%
{\mbox{}\newline\colorbox{EN}{\parbox{0.57\textwidth}%
{\small \textcolor{red}{#1}}\parbox{0.40\textwidth}%
{\footnotesize \textcolor{blue}{#2}}}\newline}

\definecolor{LM}{RGB}{68,170,153}
\definecolor{JK}{RGB}{255,255,0}
\definecolor{MH}{RGB}{204,102,119}
\definecolor{MD}{RGB}{153,153,51}
\definecolor{EN}{RGB}{136,204,238}

\DeclareMathSymbol{\de}{\mathalpha}{letters}{"40}

\usepackage{prettyref}
\newrefformat{chap}{\hyperref[#1]{Chapter~\ref*{#1}}}
\newrefformat{sec}{\hyperref[#1]{Section~\ref*{#1}}}
\newrefformat{sub}{\hyperref[#1]{Section~\ref*{#1}}}
\newrefformat{apdx}{\hyperref[#1]{Appendix~\ref*{#1}}}
\newrefformat{lst}{\hyperref[#1]{Listing~\ref*{#1}}}
\newrefformat{fig}{\hyperref[#1]{Figure~\ref*{#1}}}
\newrefformat{tab}{\hyperref[#1]{Table~\ref*{#1}}}
\newrefformat{eqn}{\hyperref[#1]{Equation~\ref*{#1}}}
\newcommand{\Ref}[1]{\prettyref{#1}}

\usepackage{url,hyperref,
microtype,subcaption}\providecommand{\linenumbers}{}

\usepackage[onehalfspacing]{setspace}

\hypersetup{colorlinks=true, breaklinks=true, linkcolor=darkblue, urlcolor=darkblue, citecolor=darkblue}

\linenumbers

\def\keyFont{\fontsize{8}{11}\helveticabold }
\def\firstAuthorLast{J. Krishnan {et~al.}} 
\def\Authors{\makebox{J. Krishnan$^{1\,2\,4}$, P.G.L. Porta~Mana$^{1}$, M.
  Helias$^{1\,2\,3}$, M. Diesmann$^{1\,2}$, E.~Di~Napoli$^{2\,4}$}}
\def\Address{$^{1}$Institute of Neuroscience and Medicine (INM-6), J\"{u}lich Research Centre, Germany\\
$^{2}$Institute for Advanced Simulation (IAS), J\"{u}lich Research Centre, Germany\\
$^{3}$Department of Physics, Faculty 1, RWTH Aachen University, Aachen, Germany\\
$^{4}$Aachen Institute for advanced study in Computational Engineering Science (AICES), RWTH Aachen University, Germany}
\def\corrAuthor{Jeyashree Krishnan}
\def\corrEmail{j.krishnan@fz-juelich.de}

\begin{document}
\onecolumn
\firstpage{1}

\title[\pdftitle]{\pdftitle}

\author[\firstAuthorLast ]{\Authors} 
\address{} 
\correspondance{} 

\extraAuth{}

\maketitle

\begin{abstract}

  \section{}
  
  Spiking neuronal networks are usually simulated with three main
  simulation schemes: the classical time-driven and event-driven schemes,
  and the more recent hybrid scheme. All three schemes evolve the state of
  a neuron through a series of checkpoints: equally spaced in the first
  scheme and determined neuron-wise by spike events in the latter two. The
  time-driven and the hybrid scheme determine whether the membrane
  potential of a neuron crosses a threshold at the end of of the time
  interval between consecutive checkpoints. Threshold crossing can,
  however, occur within the interval even if this test is negative. Spikes
  can therefore be missed.

  The present work derives, implements, and benchmarks a method for perfect
  retrospective spike detection. This method can be applied to neuron
  models with affine or linear subthreshold dynamics. The idea behind the
  method is to propagate the threshold with a time-inverted dynamics,
  testing whether the threshold crosses the neuron state to be evolved,
  rather than vice versa. Algebraically this translates into a set of
  inequalities necessary and sufficient for threshold crossing. This test
  is slower than the imperfect one, but faster than an alternative perfect
  tests based on bisection or root-finding methods. Comparison confirms
  earlier results that the imperfect test rarely misses spikes (less than a
  fraction $1/10^8$ of missed spikes) in biologically relevant settings.

  This study offers an alternative geometric point of view on neuronal dynamics.

\tiny
 \keyFont{ \section{Keywords:} State-space analysis, NEST, time-driven,
   event-driven, simulation, LIF neuron, differential geometry} 
\end{abstract}

\section{Introduction}
\label{sec:introduction}

In the last decade, considerable work has been devoted to improve the
accuracy of simulators that are capable of efficiently simulating
large networks of spiking neurons \citep{Hansel98, Mattia00,
  Shelley01, Dehaene05_0910, Morrison06c, Brette07_2604,
  DHaene-2009_1068, Elburg-2009_1913,Zheng09,Hanuschkin10_113}.  The
field is driven by the ideal of combining the capability to cope with
the high-frequency of synaptic events arriving at a neuron in nature
with a mathematically accurate implementation of the threshold process
a wide class of neuron models is based on.

Two classical schemes to simulate neuronal networks are the
time-driven and the event-driven schemes
\citep{Fujimoto00,Zeigler00,Ferscha96}. Both schemes describe the
state of the neurons by a set of variables and the action potentials
as events that mediate the interaction between them.

In a time-driven scheme, the state of a neuron is updated on a time
grid defined by the simulation step \citep[for a review
  see][]{Morrison08_267}. After all neurons are updated, their
membrane potential is checked for threshold crossings. If the membrane
potential of a neuron is above the threshold at this checkpoint, a
spike is delivered to all neurons it is connected to. Subsequently, a
new iteration step begins. The step size stipulates how frequently
occurrences of threshold crossings are inspected during the
simulation. The choice of the step size a trade-off between spike
accuracy and the speed of the simulation \cite{Morrison06c}.  Such
grid-constrained simulations force each spike event to a position on
the equidistant temporal grid spanned by the step size and therefore
induce artificial synchronization of the network dynamics
\citep{Hansel98, Shelley01, Morrison06c,
  Brette07_2604,Elburg-2009_1913,Hanuschkin10_113}.

In an event-driven scheme, the state of a neuron is updated only when
it receives a spike. A central queue of events is maintained and each
spike is inserted into this queue with its own time stamp.  Upon
update a neuron predicts when its next spike will occur in the absence
of further input. This preliminary event is inserted into the queue
and confirmed if it becomes due or removed when invalidated by further
input. Efficient and elegant predicition methods have been developed
for classes of neuron models without invertable dynamics
\citep{Brette07_2604,Elburg-2009_1913,DHaene-2009_1068,DHaene-2010_1468, Ferscha96}.
However, maintaining a central queue in a distributed simulation is
challenging and may compromise the time performance of the simulator
\citep[for a detailed review see]{Hanuschkin10_113}.

A hybrid scheme circumvents the shortcomings of both these schemes by
embedding a locally event-driven algorithm for each neuron into a globally
time-driven scheme \citep{Morrison06c}. The arrival of a spike at a
neuron introduces one additional update and check point. The dynamics
of a given neuron is then propagated from incoming spike to incoming
spike and eventually to the end point of the global timestep. If the
membrane potential of a neuron is above the threshold at a local or
global check point, the precise point of threshold crossing is determined in
continuous time, and a spike is emitted. Next to their location on the
time grid, in this scheme spike events carry a floating point offset.
Thus, in contrast to an event-driven scheme, the hybrid scheme does
not predict future spike times but identifies threshold crossings only
retrospectively. \citet{Hanuschkin10_113} demonstrate that the latter
scheme is equally accurate as the former at lower computational costs.

\begin{figure}[tbp]
  \centering
  \includegraphics[width= 0.5\textwidth]{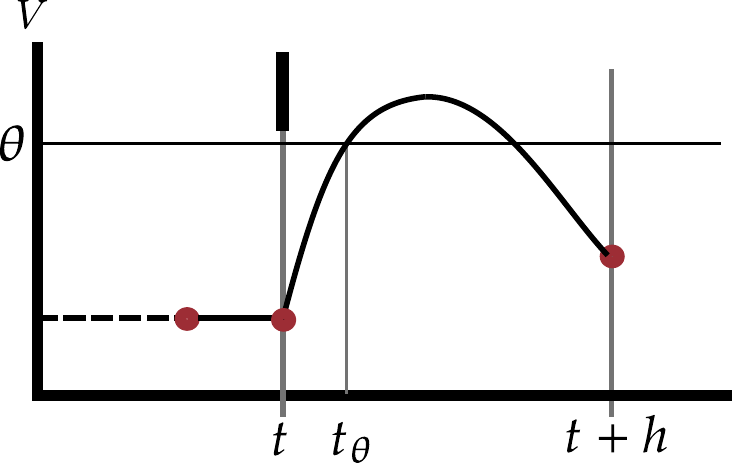}
  \caption{Illustration of undetected threshold crossing between two
    consecutive checkpoints $t$ and $t+\yh$ on the time grid. The short
    black vertical bar represents an incoming spike which causes an
    increase in the membrane voltage of the neuron, leading to a threshold
    crossing at $t_{\theta}$. The subthreshold dynamics, however, brings
    the voltage under threshold again at the next checkpoint. Since the
    test $\yV(t+\yh) \ge \yvt$ yields false, the outgoing spike at
    $t_{\theta}$ is missed. The red dots indicate points where the values
    of state variables are known.}
 \label{fig:missed-excursion}
\end{figure}

The hybrid scheme still has a loophole, however: spikes can be missed.
The reason is that, as in the time-driven scheme, crossing of the
threshold voltage is tested by inspecting whether the membrane
potential of the neuron is above threshold at the end of a checkpoint
(see \Ref{fig:missed-excursion} for an illustration of the scenario).
Nevertheless, the membrane potential $\yV$, evolved by the
subthreshold dynamics, can be below threshold $\yvt$ at two
consecutive checkpoints $t$ and $t+\yh$ while a double, quadruple,
\etc\ threshold-crossing occurred in between.  The first crossing
constitutes a missed spike. Symbolically,
\begin{equation}
  \label{eq:test_above_threshold}
  \yV(t+\yh) \ge \yvt \quad\Longrightarrow \quad\text{threshold crossing}
\end{equation}
so the test $\yV(t+\yh) \ge \yvt$ is a sufficient but not a necessary
condition for the occurrence of a threshold crossing during the time
interval $\opcl{t,t+\yh}$. For future reference we call this test
\enquote{standard test}.

The conceptual question therefore remains whether a globally
time-driven scheme can formulated such that it detects every threshold
crossing.  Although \citet{Hanuschkin10_113} argue that the loss of
spikes of the standard test is not of practical relevance in natural
parameter regimes, the availability of a method perfect by
construction would free the researcher from inquietude and costly
controls when faced with previously unexplored neuron models or
network architectures.

In this work we propose a new spike-detection method, which we term
\enquote{\sda} or \enquote{lossless test} because it is a necessary
and sufficient condition for threshold crossing to occur in a given
time interval. The method is based on state-space analysis and works
with any neuron model with affine or linear subthreshold dynamics. It
consists of a system of inequalities -- some linear, some non-linear
in the state-space variables -- that together determine whether the
initial state of a neuron will or will not reach threshold within the
time interval until the next checkpoint.
The \sda\ replaces the standard test
\eqref{eq:test_above_threshold} in the time-driven and the hybrid
scheme. Alone, it does not solve the problem of artificial
synchronization the time-driven scheme suffers from. Hence the
method is most meaningful within the hybrid scheme.
Thanks to its perfect spike detection the \sda\ can in fact be used to
benchmark the hybrid scheme based on the standard test;
\citet{Hanuschkin10_113} use the method of \citet{DHaene-2009_1068}
for this purpose.

In \Ref{sec:solution} we present the idea behind the \sda\ for a
general neuron model with an affine or linear dynamics, and develop
its mathematical construction. Parts of this construction must be
addressed on a case-by-case basis; therefore in
\Ref{sec:implementation_example} we provide a concrete implementation
of the \sda\ for the leaky integrate-and-fire model with exponential
synaptic currents \citep{Fourcaud02}, within the hybrid scheme. The
method can be algorithmically expressed in different ways.
We explore two alternative cascades of inequalities and assess their
costs in terms of time-to-completion relative to each other and to the
hybrid scheme based on the standard test
\eqref{eq:test_above_threshold}. For the latter scheme we also assess
the number of missed spikes in commonly considered network regimes.
The hybrid scheme based on the \sda\ delivers the desired exact
implementation of the mathematical definition of the neuron model
without any further approximation up to floating point precision.

Preliminary results have been published in abstract form
\citep{Kunkel_2011_P229,Krishnan16}.  The technology described in the
present article will be made available with one of next major releases
of the open-source simulation software NEST.
%
%
The conceptual and algorithmic work described here is a module in the
long-term collaborative project to provide the technology for neural
systems simulations \citep{Gewaltig_07_11204}.

\section{A time-reversed state-space analysis} 
\label{sec:solution}

\subsection{Idea: moving a surface backwards instead of a point forward} 
\label{sec:idea_explanation}

Let us summarize the problem mentioned in the previous section. We assume that a neuron's state evolves according to three different
dynamics:
\begin{enumerate*}[label=(\alph*)]\item\label{item:integrable_dyn} 
  an integrable subthreshold dynamic as long as the neuron's
  membrane potential is below threshold and there are no changes in input
  currents;
\item\label{item:jumps} discrete jumps in the subthreshold dynamics at
  predetermined times, corresponding to incoming spikes or to sudden
  changes in external currents; these can be formally incorporated into the
  subthreshold dynamics \ref{item:integrable_dyn} via delta functions; and
\item a \enquote{spike}, \ie\ an instantaneous jump of the membrane
  potential from threshold to a reset value, as soon as the potential
  reaches the threshold value. The jump may be followed by a refractory
  period in which the membrane potential remains constant at the reset
  value. Then the integrable dynamics takes place again.
\end{enumerate*}

The advantage of the integrable dynamics is that the state of the neuron at
a time $t+\yh$ can be analytically determined by that at time $t$; here
$\yh$ can be negative or positive. The evolution can thus be calculated in
discrete time steps, in particular in between times at which jumps
\ref{item:jumps} occur. The spike part of the dynamics, however, forces us
to check whether the membrane potential $\yV$ reached a threshold value
$\yvt$ within the timestep interval $\opcl{t,t+\yh}$. We call this event
\emph{threshold crossing} (by \enquote{crossing} we also mean tangency). A
sufficient condition for threshold crossing is that the membrane potential
be
above threshold at the end $t+\yh$ of the time step: by continuity, it must
have assumed the threshold value at some time in the interval
$\opcl{t,t+\yh}$. But this condition is not necessary: during the time step
the potential may touch or surpass the threshold value and then go below it
again, as in \Ref{fig:missed-excursion}, an even number of times. Its value
is then below threshold at both ends of the time step, $t$ and $t+\yh$. A
test that only relies on the sufficient condition $\yV(t+\yh) \ge \yvt$ --
the standard test~\eqref{eq:test_above_threshold} -- can therefore miss
some spikes, leading to an incorrect dynamics. We need a test based on a
necessary and sufficient condition.

\begin{figure}[!tb]
\includegraphics[width=0.9\linewidth]{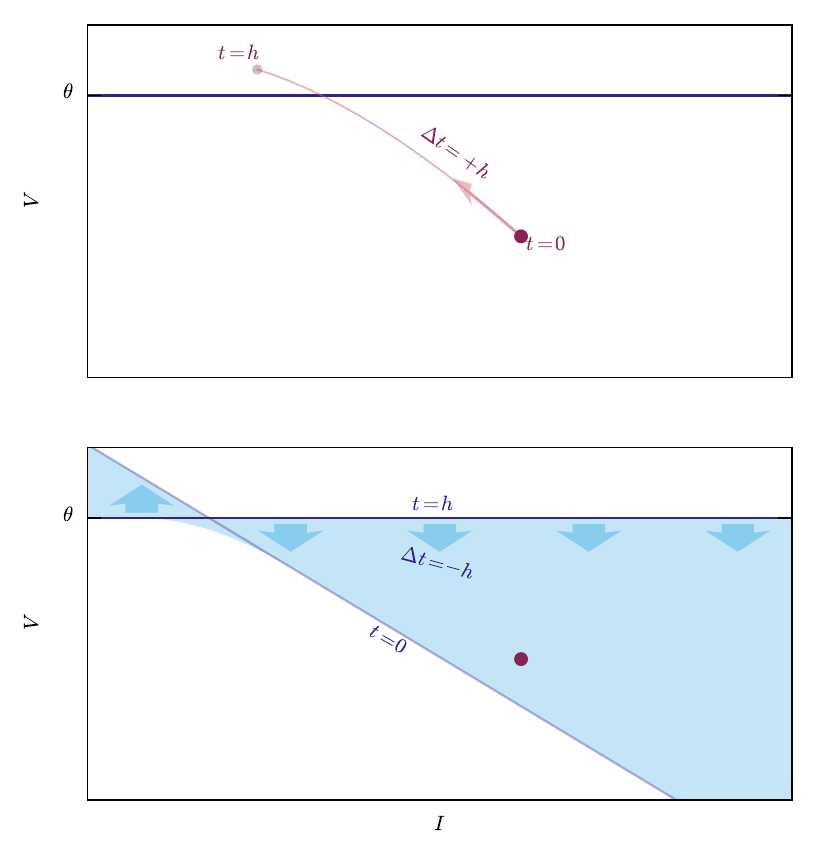}
\caption{Illustration of two exact methods to check whether an initial
  state (red dot) crosses the threshold (blue horizontal line) during the
  evolution from $t$ to $t+\yh$. \textbf{Upper panel:} the idea of a
  root-finding method is to evolve the state forwards in time by a step
  $\ydt= +\yh$, and to check whether its trajectory (light-red curve)
  intersects the threshold. Such a method informs us of where the
  intersection occurs on the threshold, and at which time. \textbf{Lower
    panel:} the idea of the \sda\ is to evolve the whole threshold
  \enquote{backwards in time} from $t=\yh$ to $t=0$ by a step $\ydt= -\yh$
  and to check whether its trajectory, which is a volume in state-space
  (light-blue region), contains the initial state, which is kept fixed. The
  threshold shifts and rotates as it evolves, and the trajectories of its
  individual points are unknown: this method does not inform us of when and
  where on the crossing it occurs, and is therefore computationally faster.}
\label{fig:idea_back_prop}
\end{figure}

A necessary and sufficient condition for threshold crossing is that the
trajectory of the state during $\opcl{t,t+\yh}$ intersect
the hypersurface
\enquote{$\text{membrane potential} = \text{threshold value}$}. Translated
into analytic geometry this means finding the solutions of a system of
parametric equations -- one representing the threshold hypersurface, the
other the trajectory -- and to test whether its solution set is empty (no
threshold crossing) or not (threshold crossing). This idea is illustrated
in \Ref{fig:idea_back_prop}A for a two-dimensional state space. This system
is usually transcendental and its solutions have to be found numerically.
Unfortunately, numerical solutions typically
rely on bisection algorithms \cite[ch.~9]{pressetal1988_r2007},
involving an increasingly finer timestepping of the dynamics. This
nullifies the advantage of having an integrable dynamics with
coarse-grained time steps.

The problem is that the test of intersection between trajectory and
threshold tells us not only whether a threshold crossing occurs, but also
the time at which it does. The test's high computational cost partly comes
from delivering this additional information. This problem is avoided if we
formulate a different geometric test that tells us whether a threshold
crossing occurs but does not deliver the crossing time. Mathematics often
offer non-constructive proofs: \enquote{there exists a solution to the
  problem, but we do not know what the solution is}. It turns out that this
ignorance is exactly what we need in our problem.

Instead of evolving the state of the neuron forwards in time, tracing a
trajectory in state-space, and checking if and when it crosses the
threshold hypersurface, we can evolve the threshold hypersurface backwards
in time, sweeping a hypervolume, and check if and when it \enquote{crosses}
the initial neuron state, which is a point. In other words we are testing
whether a point belongs to a particular state-space region. The test for
the intersection of a 1-dimensional curve with an $(\yN-1)$-dimensional
surface is replaced by the test for the membership of a point in an
$\yN$-dimensional volume. This idea is illustrated in
\Ref{fig:idea_back_prop}B for a two-dimensional state-space.

Mathematically the latter test translates into a system of inequalities
that the initial state at $t$ must satisfy if it does not cross threshold
within $\opcl{t, t+\yh}$. Even if transcendental functions appear in this
system, we do not need to find their roots: we only need to test whether
the inequalities are satisfied by simply inserting the value of the
state-space variables in the functions.

The equations corresponding to these inequalities represent the
piecewise-differentiable boundary between the set of states that will cross
the threshold within the timestep $\yh$ -- which we call \emph{\spike\
  region} -- and the set of those that won't -- which we call
\emph{\nospike\ region}. Finding these inequalities in explicit form is the
most important point of this method, and can be achieved with this
heuristic procedure:
\begin{enumerate}[label=\Roman*.,ref=\Roman*]
\item\label{item:hyperv_param} Find the hypervolume swept by the threshold,
  in parametric form. This is done by representing the backpropagation of
  the threshold in state-space as a map between two manifolds: the product
  manifold threshold${}\times{}$time, and the state-space manifold.
\item\label{item:boundary_param} Find the boundary of the hypervolume, in
  parametric form. This is done by determining the placement of the images of
  the boundaries of the product manifold and, most important, of the images
  of the critical points of the map. The latter are the points at which the
  map becomes singular, thus mapping hypervolume elements into hypersurface
  elements
  (\citealt[\chap~2]{spivak1970_r1999}; \citealt[\sect~II.A.1]{choquetbruhatetal1977_r1996}). 
\item\label{item:implicit_locus} Transform the equations of the boundary
  above from parametric to implicit form. If this is not possible, the
  boundary can still be approximated by a triangularization, as finely as we
  please.
\item\label{item:exclude_interior} Finally, exclude the parts of the
  boundary that lie in the interior of the hypervolume (so technically not
  a boundary). Some guidelines to achieve this can be given based on the
  the convex properties of the hypervolume.
\end{enumerate}
In the following analysis we mathematically develop the first two steps in
a general way for any neuron model with an affine dynamic. Let us point out
that they can be generalized to other kinds of dynamics. The procedure for
the last two steps depends on the particular neuron model, so we can only
give general guidelines. \Ref{sec:implementation_example} provides a
concrete example of all steps. An advantage of this procedure is that it
can be geometrically explained and needs to be carried out only once for
any given neuron model.

The advantage of this novel test is that its computational cost is
distributed differently from the approach where the trajectory of the
initial point is followed.
In hand-waving terms, in the trajectory-based test the crossing time is
immediately available from the parametrization of the orbit; the
threshold-crossing condition is represented by computationally costly
inequalities, because the crossing time, when it exists, can be read from
them immediately. In the novel test the crossing time is a more complicated
function of the hypervolume parametrization; the threshold-crossing
condition in this case is represented by inequalities that are
computationally less costly, because they do not explicitly deliver the
crossing time. Knowledge of the latter requires an additional computational
cost. The novel test thus allows us to avoid this extra cost when we check
for the existence of a threshold-crossing; we only need to pay it if the
crossing does occur.

From a mathematical point of view the two approaches are equivalent. The state-space technique is just a useful reformulation of
the conventional trajectory view that is obtained by suitable
mathematical manipulations. The conceptual device of propagating the
threshold backwards in time is useful because it performs those
manipulations in a transparent manner and gives them an intuitive
dynamical meaning.

\subsection{Mathematical preliminaries}
\label{sec:maths_prelim}

The final equations to be obtained,
\eqn~\eqref{eq:nospike_final_inequalities}, can be derived by concepts
from vector analysis, Cartesian geometry, and functional analysis; but
the derivation is lengthy. To shorten it we use concepts and
terminology from affine spaces
\citep{coxeter1961_r1969,rockafellar1970_r1972,nomizuetal1994,artin1955,portamana2011},
and differential manifolds
\citep{choquetbruhatetal1977_r1996,burke1985_r1987,burke1995,simonetal1992,nomizuetal1994,marsdenetal1983_r2007,bossavit1991,bossavit1994_r2002,schouten1951_r1989,derham1955_t1984,dodsonetal1977_r1991,ramanan2005}.

The state-space $\yM$ of a neuron has a natural vector-space structure
(an affine-space structure would also suffice), inherited from the
physical quantities that define it: the membrane potential $\yV$ and
other $\yN-1$ physical quantities $\yI$ whose exact number and
definition depend on the specific neuron model (\eg, $\yI$ could
represent currents or additional voltages, for example of different
compartments). For our purpose it is useful to consider $\yM$ as an
$\yN$-dimensional differential manifold: its points $\set{\ys}$ are
the neuron states, and the quantities $(\yI, \yV)$ are coordinates
$\yV\colon \yM \to \RR$ and $\yI \colon \yM \to \RR^{\yN-1}$ \eg
, the membrane potential of a state $\ys$ is $\yV(\ys)$. These
coordinates respect the vector structure of the state-space,
\ie\ $\yV(\ys_1 + \ys_2) = \yV(\ys_1) + \yV(\ys_2)$ and likewise for
$\yI$.

Every hyperplane in the state-space is defined by an affine equation
$\ygr\ys = -\ygs$, the covector $\ygr$ being the normal to the hyperplane,
and $-\ygs$ being the affine term. The inequality $\ygr\ys > -\ygs$ defines one
of the two half-spaces delimited by the hyperplane. The threshold
hyperplane is especially important: it is the set of states $\ys$ whose
membrane potential has the threshold value:
\begin{equation}
  \yV(\ys)=\yvt.
  \label{eq:threshold_hyperplane}
\end{equation}
Its equation $\ygr\ys = \ygs$ in coordinates $(\yI, \yV)$ has coefficients
\begin{equation}
  \label{eq:affine_form_threshold}
  \begin{gathered}
  \ygr=(\yze\T, 1), \qquad \ygs=\yvt,
\\
\ygr\ys = \ygs\quad\text{on threshold,}
\qquad
\ygr\ys <\ygs\quad\text{below threshold.}
\end{gathered}
\end{equation}

An affine transformation of the state-space onto itself,
\begin{equation}
  \label{eq:affine_transf}
  \ys \mapsto \ygM \ys + \ygv,
\end{equation}
where $\ygM$ is a linear transformation and $\ygv$ a state, maps each
hypersurface and half-space $\ygr\ys \ge \ygs$ to a hypersurface and
half-space $\ygrb\ys \ge \ygsb$ with
\begin{equation}
  \label{eq:new_hypersurface}
  \ygrb = \ygr\ygM^{-1}, \qquad
  \ygsb =  \ygs + \ygr\ygM^{-1}\ygv
\end{equation}
(the transformation of the normal $\ygr$ shows why it is a covector rather
than a vector).

We now show that the integrable part of the neuron dyamics within a
finite time step $h$ is an affine transform.  Similarly, the integrable part of
of the neuron dynamics we consider an affine evolution determined by
the equation
\begin{equation}
  \label{eq:linear_evolution}
  \ysd(t) = \yA \ys(t) + \ym.
\end{equation}
In a time interval $\yh$, this dynamics propagates an initial state $\yso$
at time $\yto$ into the final state
\begin{equation}
  \label{eq:evolution_initial_state}
  \ys(\yto+\yh) = \e^{\yh\yA}\ys(\yto) + \bigl(\e^{\yh\yA}-1\bigr)\yA^{-1}\ym,
\quad \ys(\yto)=\yso.
\end{equation}
This, for each $\yh$, is an affine transformation of the
form~\eqref{eq:affine_transf}. In coordinates $(\yI, \yV)$ the linear
operator $\yA$ and vector $\ym$ have the block form
\begin{equation}
  \label{eq:A_m_blockform}
 \yA =
 \begin{pmatrix}
   \yB & \yac \\ \yar & \yal
 \end{pmatrix},
\qquad
\ym =
\begin{pmatrix}
  \yrc \\ \ymu
\end{pmatrix},
\end{equation}
where $\yB$ is an $(\yN-1,\yN-1)$ matrix, $\yal$ and $\ymu$ numbers, and the
dimensionalities of the rectangular matrices $\yar$, $\yac$ follow accordingly.

\subsection{Derivation of the threshold-crossing condition}
\label{sec:derivation_thresholdcrossing}

Let us mathematically summarize the first threshold-crossing test discussed
in \Ref{sec:idea_explanation}. We said that the state evolution
\eqref{eq:evolution_initial_state} can be efficiently used in a time-step
scheme in numerical simulations, but we need to test whether a threshold
crossing occurred at some time $t \in \opcl{\yto,\yto+\yh}$. A necessary
and sufficient condition would be the existence of solutions of the
trascendental equation in $t$
\begin{equation}
  \label{eq:threshold_crossing_t}
  \yV\bigl[\e^{t\yA}\ys(\yto) +
  \bigl(\e^{t\yA}-1\bigr)\yA^{-1}\ym\bigr]
  = \yvt,
  \qquad t \in \opcl{0,\yh},
\end{equation}
which corresponds to the intersection of the
trajectory~\eqref{eq:evolution_initial_state} and the threshold
hyperplane~\eqref{eq:threshold_hyperplane}; but it is a costly condition to
test.

We now develop the second kind of threshold-crossing test discussed in
\Ref{sec:idea_explanation}, according to the
steps~\ref{item:hyperv_param}--\ref{item:exclude_interior}.
Steps~\ref{item:hyperv_param} and~\ref{item:boundary_param} are performed
in full generality for an affine dynamics. Steps~\ref{item:implicit_locus}
and~\ref{item:exclude_interior} have to be solved on a case-by-case basis,
so their analysis below is only a guideline; a concrete example on how to
perform them is given in \Ref{sec:boundary_implicit_example}
and \Ref{sec:boundary_intersections_example}.

\subsubsection{Hypervolume in parametric form}
\label{sec:hyperv_parametric}

Consider the threshold hyperplane $\yV(\ys)=\yvt$ as an
$(\yN-1)$-dimensional manifold with coordinates $\yu$. It is embedded in
state-space via the map
\begin{equation}
  \label{eq:threshold_manifold}
  \yu \mapsto (\yI, \yV) = (\yu,\yvt).
\end{equation}
Each state $(\yu, \yvt)$ on the threshold, when propagated backwards in
time for an interval $\yh$, traces a curve in state-space (the yellow lines
of \Ref{fig:bounding_surface_jacobian}).

The union of these curves is an
$\yN$-dimensional product manifold, called the \emph{extrusion}
\citep{bossavit2003b} of the threshold hyperplane. We can use coordinates
$(\yu,t)\in (\RR^{\yN-1}\times \clcl{0,\yh})$ on this manifold. Its mapping
into state-space is given, with the help of
\eqn~\eqref{eq:evolution_initial_state}, by
\begin{equation}
  \label{eq:threshold_manifold_backpropagated}
 \ybt\colon (\yu,t) \mapsto \e^{-t\yA} \begin{pmatrix}
    \yu\\ \yvt
  \end{pmatrix} +\bigl(\e^{-t\yA}-1\bigr)\yA^{-1}
  \begin{pmatrix}
    \yrc \\\ymu
  \end{pmatrix},
\end{equation}
where $t>0$ is the direction of the past. This map is analytic, but
generally not an \emph{embedding} because it can have self-intersections. We will
see the significance of this in \Ref{sec:region_missed_spikes}.
It is not an \emph{immersion} either because it can have singular points;
these will be especially important for us because they constitute part of
the boundary between \spike\ and \nospike\ regions. See
\Ref{fig:bounding_surface_jacobian} for a two-dimensional example.

For fixed $t$, the map $\yu\mapsto\ybt(\yu,t)$ is affine, and its image is
a hyperplane representing the states on the threshold propagated backwards
in time for an interval $t$. Combining it with
\eqns~\eqref{eq:affine_form_threshold}--\eqref{eq:new_hypersurface} we find
that the backpropagated threshold has $t$-dependent normal and affine terms
\begin{equation}
  \label{eq:equation_plane_backpropagated}
\begin{split}
\ygr_t &=
\begin{pmatrix}
  \yze\T & 1
\end{pmatrix}
\e^{t\yA},
\\
\ygs_t &= \yvt
+
\begin{pmatrix}
  \yze\T & 1
\end{pmatrix}
\bigl(1- \e^{t\yA}\bigr)\yA^{-1}
  \begin{pmatrix}
    \yrc \\\ymu
  \end{pmatrix}.
\end{split}
\end{equation}
The inequality $\ygr_t \ys < \ygs_t$ determines the backpropagated
half-space, which is below threshold.

\subsubsection{Hypervolume boundary in parametric form}
\label{sec:hyperv_boundary_parametric}

We must now find the boundary of the image of the map $\ybt$. The latter is
a closed set, being the image of a closed set under a continuous map; its
boundary must therefore be the image of some points of the domain. Such
points must either lie on the boundaries of the domain,
$\RR^{\yN-1} \times \set{0}$ and $\RR^{\yN-1} \times \set{\yh}$, or be
critical points of $\ybt$, or both, because $\ybt$ is differentiable. See
the example of \Ref{fig:bounding_surface_jacobian}.

The images of the boundary are easily found from
\eqref{eq:threshold_manifold_backpropagated}: one (image of $t=0$) is the
threshold hyperplane, the other ($t=\yh$) is the hyperplane
$\ygr_{\yh} \ys = \ygs_{\yh}$, with coefficients given by
\eqref{eq:equation_plane_backpropagated}. Explicitly, in coordinates
$(\yI, \yV)$,
\begin{align}
  \label{eq:boundary_t0}
&\yV = \yvt,
  \\
  \label{eq:boundary_th}
  &\begin{pmatrix}
  \yze\T & 1
\end{pmatrix}
\e^{\yh\yA}
\begin{pmatrix}
             \yI \\ \yV
           \end{pmatrix}
  =
 \yvt
+
\begin{pmatrix}
  \yze\T & 1
\end{pmatrix}
\bigl(1- \e^{\yh\yA}\bigr)\yA^{-1}
  \begin{pmatrix}
    \yrc \\\ymu
  \end{pmatrix}.
\end{align}

Let us find the image of the critical points of $\ybt$. The tangent map of
$\ybt$ at a point $(\yu,t)$ is
\begin{equation}
  \label{eq:tangent_map}
\Tg\ybt(\yu,t) = (\de_{\yu}\ybt, \de_t\ybt) =
- \e^{-t\yA}
\begin{pmatrix}
\idm & \yB\yu + \yac \yvt + \yrc \\
\yze\T &\yar\yu + \yal\yvt + \mu 
  \end{pmatrix}.
\end{equation}
The tangent map is also denoted $\ybt_*$ in some differential-geometry
texts.

Its determinant is the inverse ratio between a volume element at that point
and its image in state-space (the sign determines their relative
orientation). Hence this ratio vanishes at points where volume elements are
mapped onto area elements (in other words, $\yN$ linearly independent
vectors in the domain are mapped onto $\yN$ linearly dependent vectors),
which is a feature of the boundary. See the two-dimensional example of
\Ref{fig:bounding_surface_jacobian}.

Let us look for points where $\det\Tg\ybt(\yu,t) = 0$. In
\eqn~\eqref{eq:tangent_map}, the determinant of the exponential never
vanishes, so we only have to consider the determinant of the matrix on the
right. This is easily calculated by Laplace expansion along the last row,
whose elements all vanish except the last. The cofactor of the last element
is $\det\idm$ (modulo a sign). Hence
\begin{equation}
  \label{eq:determinant_eqn_explicit}
  \det\Tg\ybt(\yu,t) = 0 \quad\Longleftrightarrow\quad \yar\yu + \yal\yvt + \ymu = 0
\end{equation}
and the coordinates $(\yu,t)$ of critical points satisfy
\begin{equation}
  \label{eq:boundary_implicit}
  \yar\yu + \yal\yvt + \ymu = 0,\qquad 0\le t\le \yh.
\end{equation}
This equation says that one of the coordinates $\yu$ has an affine
dependence on the remaining ones; let us call these $\yv$. For example, if
$\yarr_{\yN-1} \ne 0$, the equation above has the parametric solution
\begin{equation}
  \begin{gathered}
    \yuu_1 = \yvv_1, \qquad \dotso,\qquad \yuu_{\yN-2} = \yvv_{\yN-2},
    \\
    \yuu_{\yN-1}=
    -\frac{\yarr_1\yvv_1 + \dotsb  + \yarr_{\yN-2} \yvv_{\yN-2} +\ymu+\yal\yvt}{\yarr_{\yN-1}}.
  \end{gathered}
\label{eq:example_one_less_coord}
\end{equation}
Denote this affine dependence by $\yu(\yv)$. By taking the derivative of
\eqn~\eqref{eq:boundary_implicit} with respect to $\yv$ we have
\begin{equation}
  \label{eq:zero_product}
  \yar\de_{\yv}\yu = \yze\T,
\end{equation}
a property we will use later.

The locus of critical points in state-space is then given parametrically by
a map $\ybo$, found by substitution of
\eqn~\eqref{eq:example_one_less_coord} in
\eqref{eq:threshold_manifold_backpropagated}:
\begin{equation}
  \label{eq:equation_boundary_implicit}
 \ybo\colon (\yv,t) \mapsto \e^{-t\yA} \begin{pmatrix}
    \yu(\yv)\\ \yvt
  \end{pmatrix} +\bigl(\e^{-t\yA}-1\bigr)\yA^{-1}
  \begin{pmatrix}
    \yrc \\\ymu
  \end{pmatrix},
\quad 0 < t < \yh.
\end{equation}

This locus has four important interrelated features:

First: from the form of \eqns~\eqref{eq:boundary_implicit}
and~\eqref{eq:equation_boundary_implicit}, the locus of critical points is
flat along $\yN-2$ dimensions -- corresponding to a fixed value of the
coordinate $t$ -- and is curved normally to the direction $t$: \emph{it is
  an $(\yN-2)$-ruled surface}.

Second: the points on the locus
corresponding to fixed $t$ belong to a hyperplane with
coefficients~\eqref{eq:equation_plane_backpropagated}, as is easily checked
by substitution. In other words, the locus of critical points is the
envelope of the backpropagated threshold hyperplanes $\ygr_t\ys=\ygs_t$,
\eqn~\eqref{eq:equation_plane_backpropagated}, at different times $t$, and
its tangent hyperplanes have normals $\ygr_t$ given
by~\eqref{eq:equation_plane_backpropagated}.

Third: considering the feature above for $t=0$, the locus of critical
points is tangent to the threshold hyperplane.

Fourth: comparing the dynamics~\eqref{eq:linear_evolution}, the map for the
threshold hyperplane~\eqref{eq:threshold_manifold}, and the critical-point
condition~\eqref{eq:boundary_implicit}, we notice that the latter is also
the condition for the trajectory~\eqref{eq:evolution_initial_state} of a
state $\yso=(\yu, \yvt)$ on the threshold hyperplane to have an extremum in
the membrane potential $\yV$. Hence, \emph{the locus of critical points is
  the trajectory of the intersection between the threshold and the
  $\yV$-nullcline}.

In view of the second property above, let us call the locus of critical
points \emph{\env}, for brevity.

\subsubsection{Hypervolume boundary in implicit form}
\label{sec:hyperv_boundary_implicit}

The next step is the elimination of the parameters $(\yv,t)$ to express the
\env\ $\ybo$ as one or more implicit equations
$\yS(\ys)=0$ defined on domains $\yD$ for the state-space coordinates
$\ys$. This step can involve a transcendental equation, therefore we cannot
give a general solution for it. In this regard, a criticism might be raised, which we immediately
address: it looks like we have already taken great care in avoiding to solve
\eqn~\eqref{eq:threshold_crossing_t}, only to face another difficult
transcendental equation? The current problem is more manageable for
three reasons. First, we observe that the equations needed to
re-express the curved surface in explicit form are generally easier to
handle than
\eqref{eq:threshold_crossing_t}. \Ref{sec:boundary_implicit_example}
gives a concrete example. Second, the problem of finding the explicit
form of the curved surface must be solved only once, whereas the
solution of the original equation \eqref{eq:threshold_crossing_t} has
to be found for each initial state $\yso$. Third, the equation for the
curved surface is easier to approximate numerically than the original
one \eqref{eq:threshold_crossing_t}, again because the approximations
do not depend on an initial state.

Suppose we have found a function $\yS$ such that $\yS(\ys)=0$ is the
envelope:
\begin{equation}
  \label{eq:curved_surf_implicit_equality}
  \set{\ys \st \yS(\ys) = 0, \ys \in \yD} \mathrel{\ =\ } \set{\ybo(\yv,t)
    \st \yv\in \RR^{\yN-2}, 0 \le t \le \yh}.
\end{equation}
The condition above leaves $\yS$ completely undetermined (apart from
smoothness requirements) outside of the \env. Therefore, if possible,
it is useful to extend the condition as follows. Since $\yS(\ys)=0$ is
the envelope of the backpropagated threshold hyperplanes
$\ygr_t\ys=\ygs_t$, \eqn~\eqref{eq:equation_plane_backpropagated}, at
different times $t$, it must be tangent to each of them. The
inequalities $\ygr_t\ys-\ygs_t<0$, for each $t$, correspond to the
backpropagated below-threshold half-plane, and it is useful to choose
$\yS$ in such a way that the inequality $\yS(\ys)<0$, $\ys \in \yD$,
determines the side corresponding to the intersection of these
half-spaces. An example is given in
\Ref{sec:boundary_implicit_example}. The differential $\di\yS$ then is
a positive multiple of $\ygr_t$ (note that $\di\yS$ and $\ygr_t$ are
collinear because of the tangency condition).

For the following, let us assume that $\yS$ has been chosen this way.

\subsubsection{Boundary intersections and final system}
\label{sec:boundary_intersections}

If a state does not cross the threshold hyperplane at any time $t \in
\clcl{0,\yh}$, then it must, by definition, belong for every $t$ to
the image of the half-space that lies below threshold, when this
half-space is backpropagated by the time $t$. Mathematically, this is
simply the statement of the equivalence
\begin{equation}
  \label{eq:equivalence_to-back_prop}
  \ygr(\ygM \ys + \ygv) < \ygs
  \Liff
   (\ygr\ygM) \ys <  (\ygs - \ygr\ygv)
\end{equation}
established in \Ref{sec:maths_prelim}, where $\ygM$ and $\ygv$ are
the coefficients of the affine evolution by time $t$.

By construction in the previous subsection, the inequality $\yS(\yI,
\yV) < 0$ defines the region of intersection of all such
backpropagated half-spaces, bounded by the \env\ $\yS(\yI, \yV)=0$. We
just need to join to it the condition for the boundaries corresponding
to the times $t=0$ and $t=\yh$, $\yV<\yvt$, $\ygr_{\yh} \ys <
\ygs_{\yh}$, discussed in \Ref{sec:hyperv_parametric}.

The states that do not cross the threshold during the time interval
$\clcl{0,\yh}$ belong therefore to  the \nospike\ region defined, in
coordinates, by
\begin{shaded}
  \begin{subequations}
      \label{eq:nospike_final_inequalities}
    \begin{align}
      \intertext{$\yso= (\yI,\yV) \in \text{\nospike\ region} \Liff{}$}
        &\yV <\yvt
        \\
        \intertext{and}
        &\!\begin{pmatrix}
          \yze\T & 1
        \end{pmatrix}
        \e^{\yh\yA}
        \begin{pmatrix}
          \yI \\ \yV
        \end{pmatrix}
        <
        \yvt
        +
        \begin{pmatrix}
          \yze\T & 1
        \end{pmatrix}
        \bigl(1- \e^{\yh\yA}\bigr)\yA^{-1}
        \begin{pmatrix}
          \yrc \\\ymu
        \end{pmatrix}
        \\
        \intertext{and} &\yS(\yI, \yV) < 0, \quad (\yI, \yV) \in \yD.
    \end{align}
  \end{subequations}
\end{shaded}
\noindent Some inequalities in this system may turn out to be redundant, \ie\
automatically satisfied if the remaining ones are, and can thus be dropped.
\Ref{sec:boundary_implicit_example} illustrates such a redundancy.

\subsection{The region of missed spikes}
\label{sec:region_missed_spikes}

In the previous section we mentioned that the $\yN$-dimensional product
manifold formed by the $(\yN-1)$-dimensional threshold hyperplane and the
time interval $\clcl{0,\yh}$ presents self-intersections when mapped to the
state-space. Two-dimensional examples of such a region are shown in
\Ref{fig:bounding_surface_jacobian}, lower panel, and
in \Ref{fig:spiking-region}, the region called $\Sp_2$.

A state $\ys$ in the self-intersection region corresponds to two or more
different coordinates $(\yu,t)$:
\begin{equation}
  \label{eq:two_diff_coords}
  \ys=\ybt(\yu,t) =\ybt(\yu',t'),\qquad t\ne t'.
\end{equation}
Note that $\yu\ne\yu'$, $t=t'$ is impossible for an affine
transformation, since the first column of \eqref{eq:tangent_map} can
never vanish. The condition above simply means that $\ys$ crosses the
threshold at $\yu$ after an interval $t$ and at $\yu'$ after an
interval $t'$. By continuity of the
dynamics~\eqref{eq:linear_evolution}, one of the two must be a
crossing from above, and one from below: this is exactly the scenario
of double threshold crossing illustrated in \Ref{fig:missed-excursion}.

Suppose we have a probability distribution for the initial states, for
example one that is invariant under the
dynamics~\eqref{eq:linear_evolution}. The probability of the
self-intersection region $P(\ySi)$ is the probability that the initial
state will lead to a double-crossing of the threshold, and therefore be
missed. This fact will be used in \Ref{sec:occupation_freqs} to estimate
the number of spikes missed.

\subsection{Convexity, approximations, optimization}
\label{sec:convexity_approx}

The affine dynamics~\eqref{eq:evolution_initial_state} preserves affine
combinations of solutions -- and therefore convex combinations as well. If
$\ys_1$, $\ys_2$ are two arbitrary initial states in the \nospike\ region,
then their propagated states also satisfy $\yV[\ys_1(t)]<\yvt$ and
$\yV[\ys_2(t)]<\yvt$ when $0\le t \le \yh$; and also the propagation of
their convex combination $\ylam \ys_1+(1-\ylam) \ys_2$, $0 <\ylam <1$
satisfies
\begin{equation}
  \label{eq:convex_combin_below_threshold}
\yV[\ylam \ys_1(t)+(1-\ylam) \ys_2(t)] <\yvt,
\qquad
0 <\ylam <1,
\end{equation}
\ie\ it lies in the \nospike\ region. This proves that the \nospike\ region
is \emph{convex}.

Convexity is important for approximations. If transforming
the parametric equation for the \env~\eqref{eq:equation_boundary_implicit}
into an implicit form~\eqref{eq:curved_surf_implicit_equality} turns out to
be analytically impossible, we can numerically find some points on the
\env\ and then approximate the latter by simplices constructed on these
points, as finely as needed. In other words we can triangularize the \env.
Owing to convexity, the triangularization can be done completely on the
side of the \nospike\ region (the corners of the simplices touch the \env),
or completely from that of the \spike\ region (the barycentres of the
simplices touch the \env). This way we can formulate a test with no false
positives, or one with no false negatives, or both.

The \env\ is moreover a ruled surface, as shown in
\Ref{sec:hyperv_boundary_parametric}. The triangulation on the side of the
\nospike\ region can therefore be conveniently chosen in such a way that
one face of each simplex fully lies on the \env.

\section{Implementation example:
  leaky integrate-and-fire
neuron with exponentially decaying post-synaptic currents}
\label{sec:implementation_example}
\subsection{The example model}
\label{sec:example_model}

In the previous section we mathematically developed the idea of propagating
the threshold backward in time in order to check whether a
threshold-crossing occurs in a time-stepped dynamics. The derivation, valid
for an affine subthreshold dynamics, is general and therefore also quite
abstract; moreover, it involves a couple of mathematical steps
(\ref{item:implicit_locus} and \ref{item:exclude_interior} in
\Ref{sec:idea_explanation}) for which no general formulae can be given.

To explain the idea in more concreteness and to give an example of how to
face all its steps, we now apply the scheme to a simple but relevant model
with a 2-dimensional state-space: the leaky integrate-and-fire neuron with
exponentially decaying \psc s. This model has a homogeneous linear dynamics
on a 3-dimensional state-space \citep{Rotter99a}, where the third
coordinate is the input current. If this current is constant, the dynamics
can be rewritten as a 2-dimensional affine one.

Leaky integrate-and-fire models, despite their simplicity, approximate the
behavior of real neurons with high accuracy \citep{Rauch03}. The model with
exponential synaptic currents captures important properties of real
neurons: The postsynaptic potential has a finite rise and decay time and
the membrane potential is a continuous function of time. Continuity avoids
artificial synchronization, present in simpler models. Moreover, the model
is to some extent analytically tractable. For short synaptic time
constants, the mean firing rate \citep{Fourcaud02} as well as the linear
response to small inputs \citep{schueckeretal2015} can be
obtained analytically.

\subsection{Mathematical preliminaries: terms in block form}
\label{sec:terms_block_form}

The example model has a 2-dimensional state-space for a single neuron,
defined by the post-synaptic current $\Is$ and the membrane potential
$\yV$, which are also our coordinates. Its subthreshold interspike dynamics
\ref{item:integrable_dyn} in \Ref{sec:solution} is affine:

\begin{subequations}
  \label{eq:dynamics_2D_model}
  \begin{align}
    \begin{split} 
      \Dot{\Is}(t) &= -\frac{1}{\ts} \Is(t), \\
      \Dot{\yV}(t) &= \frac{1}{\Cm}\Is(t) -\frac{1}{\tm} \yV(t) +
      \frac{1}{\Cm} \Ix,
    \end{split}
    \\
    \intertext{or in matrix form}
    \frac{\di}{\di t}
    \begin{pmatrix}
      \Is \\ \yV
    \end{pmatrix}
                   &=
                     \begin{pmatrix}
                       -\frac{1}{\ts} & 0 \\
                       \frac{1}{\Cm} & -\frac{1}{\tm}
                     \end{pmatrix}
                                       \begin{pmatrix}
                                         \Is \\ \yV
                                       \end{pmatrix}
    +
    \begin{pmatrix}
      0 \\ \frac{1}{\Cm} \Ix
    \end{pmatrix},
  \end{align}
\end{subequations}
where $\Cm$ is the membrane capacitance, $\tm$ is the membrane time
constant, $\Is$ is the synaptic input current and $\Ix$ is the external
input current. The membrane potential $\yV$ is subject to dissipation with
time constant $\tm$ and integrates the \psc\ $\Is$. The latter decays
exponentially with time constant $\ts$.  Typical values of the parameters
are
$\tm = 10 \ums$, $\Cm = 250 \upF$, $\ts= 2 \ums$, and the threshold
$\yvt = 20 \umV$.

Incoming spikes are incorporated in the equation for the current as
Dirac deltas, and the external current $\Ix$ has jump discontinuities
in time. In the timestepped evolution, such discontinuous events are
implemented as instantaneous changes in the initial state $\yso$ at
each timestep. Hence we do not need to consider them explicitly in the
equations above \citep{Rotter99a}.

In terms of the block form of \Ref{sec:maths_prelim} we have
\begin{equation}
  \label{eq:2D-case_correspondence}
  \begin{aligned}
  \yB &= 
  \begin{psmallmatrix}
    -\frac{1}{\ts}
  \end{psmallmatrix},& \yac &= 
\begin{psmallmatrix}
  0
\end{psmallmatrix}, &&& \yrc &= 
\begin{psmallmatrix}
  0
\end{psmallmatrix},
\\
\yar &= 
\begin{psmallmatrix}
  \frac{1}{\Cm}
\end{psmallmatrix}, & \yal &= -1/\tm, &&& \ymu &= \frac{1}{\Cm}\Ix.
\end{aligned}
\end{equation}
The exponential of $\yA$ is
\begin{equation}
  \label{eq:exponential_A}
  \exp(-t\yA) = 
  \begin{pmatrix}
 \etsp & 0 \\
 \frac{\bigl(\etmp-\etsp\bigr) \tm \ts}{\Cm
   (\tm-\ts)} 
& \etmp
  \end{pmatrix},
\end{equation}
which determines the evolution of the neuron state $\yso$ by an affine map as in
\eqn~\eqref{eq:evolution_initial_state}.

Since the state space is 2-dimensional, the
\enquote{hyperplanes} and \enquote{hypersurfaces} of
\Ref{sec:solution} are straight lines and curves. In
particular, the threshold hyperplane is a line; when propagated by a time
$t$ it maps onto a line with covector and affine term given by
\eqn~\eqref{eq:equation_plane_backpropagated}, explicitly
\begin{equation}
  \label{eq:propag_threshold_2D}
  \begin{aligned}
    \ygr_t &= \begin{pmatrix}
 \frac{\bigl(\etmm-\etsm\bigr) \tm \ts}{\Cm
   (\tm-\ts)} 
& \etmm
  \end{pmatrix},
\\
\ygs_t &= \yvt
-\frac{\Ix\tm \bigl(1-\etmm\bigr)}{\Cm}.
  \end{aligned}
\end{equation}

For this model, testing for threshold-crossing by checking the
intersection of a propagated state and the threshold line means
solving the following transcendental system in $t$, given the initial
state $\yso=(\Is,\yV)$:
\begin{equation}
  \label{eq:threshold_cross_time_example}
\left\{  \begin{aligned}
    &
\etmm\yV +
  \frac{\Ix\tm}{\Cm} (1-\etmm)
  +\Is\frac{\tm\ts}{\Cm}
  \frac{\etmm-\etsm}{\tm-\ts}
=\yvt,
\\
&0 \le t \le \yh,
\end{aligned}
\right.
\end{equation}
for which we cannot find a solution in analytic form; we would have to
resort to bisection algorithms.

\subsection{The threshold-crossing condition}
\label{sec:example_thresholdcrossing}

\subsubsection{Hypervolume in parametric form}

\begin{figure}[!ht]
\centering
\includegraphics[width=\linewidth]{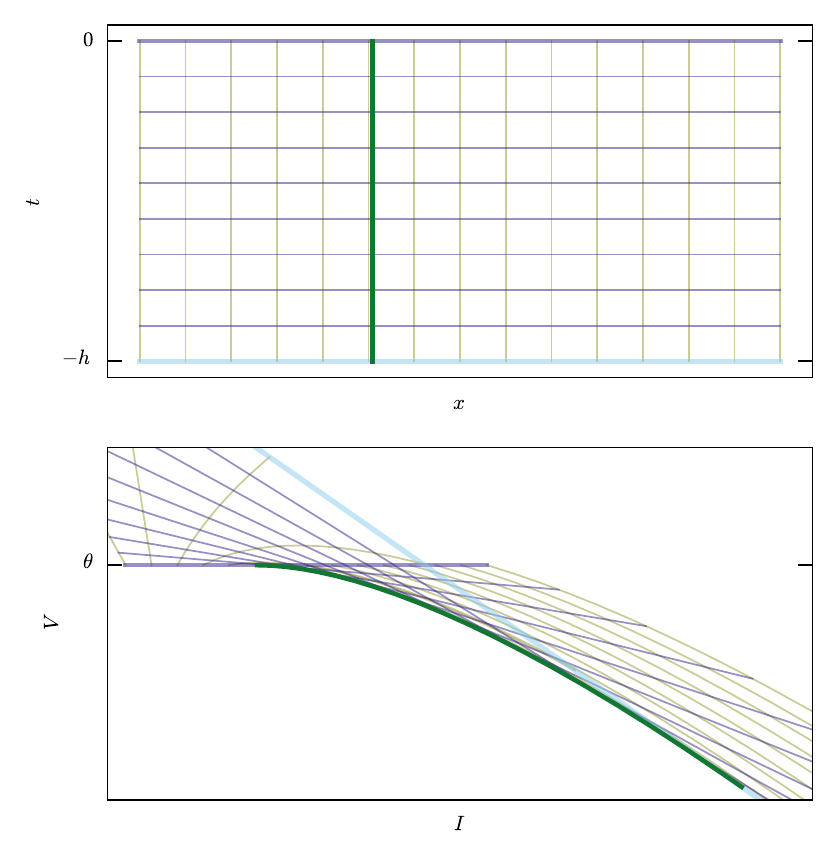}%
\caption{Example of map $\ybt\colon (\yu,t)\mapsto (\yI,\yV)$,
  \eqn~\eqref{eq:threshold_manifold_backpropagated}, in two dimensions.
  \textbf{Upper panel:} the abstract manifold with coordinates
  $(\yu,t)$ 
  corresponding to the values of current and time at threshold crossing.
  \textbf{Lower panel:} the image of the map 
  in state space.   Thick green curve corresponds to the set of singular
  points where $\mathrm{det}(TE)=0$, \eqn~\eqref{eq:threshold_crossing_t}. Yellow lines,
  constant $\yu$, are trajectories of states ending on the threshold. Violet
  lines, constant $t$, are snapshots of the threshold moving
  \enquote{backwards in time}. The thicker violet and
  blue lines correspond to the boundaries $t=0$ and $t=-\yh$.
  For $t=0$ we have $I=x$, $\yV=\yvt$.
}
\label{fig:bounding_surface_jacobian}
\end{figure}

The product manifold \enquote{threshold line${}\times{}$time interval} is
in this case 2-dimensional, with coordinates $(\yuu, t)$.
Here $\yu$ is the value of the current at threshold crossing
and $t$ the corresponding time point.
Its mapping
$\ybt$, \eqn~\eqref{eq:threshold_manifold_backpropagated}, to
the state-space is
\begin{equation}
  \label{eq:threshold_manifold_backpropagated_explicit}
  \ybt\colon (\yuu,t) \mapsto (\Is,\yV)=
  \biggl(
  \etsm \yuu,\quad
  \etmm\yvt +
  \frac{\Ix\tm}{\Cm} (1-\etmm)
  +\yuu\frac{\tm\ts}{\Cm}
  \frac{\etmm-\etsm}{\tm-\ts}
\biggr).
\end{equation}
This map is shown in \Ref{fig:bounding_surface_jacobian}: the $\yuu$
isocurves are yellow and in the lower panel they represent the trajectories
of states terminating on the threshold; the $t$ isolines are blue
and represent the threshold line propagated at different times.
Each area element $\di\yu\land\di t$ in
the domain -- the small rectangles in the upper plane -- is mapped into an
area element $\di\yI\land\di\yV$ in the image. Note how these area elements
are rotated and sheared. The thicker green curve is the set of singular
points where the determinant of the tangent map vanishes:
$\det(\Tg\ybt)=0$. Such points are singular because around them the images
of the area elements get flattened to one dimension.
In \ref{sec:region_missed_spikes_2D} we discuss the region of
self-intersection bounded by the thick light cyan line, the thick dark blue
line, and the thick green curved line.

\subsubsection{Hypervolume boundary in parametric form}

The boundaries of the image of the map $\ybt$ must be, as explained in
\Ref{sec:hyperv_boundary_implicit}, a subset of the images of the
boundaries of the domain, $\RR\times\set{0}$ and $\RR\times\set{\yh}$, and
of the \env.

The images of the boundaries, with general equations
\eqref{eq:boundary_t0} and \eqref{eq:boundary_th}, in terms of
$(\Is,\yV)$ follow
\begin{align}
  \label{eq:boundary_t0_example}
&\yV = \yvt,
  \\
  \label{eq:boundary_th_example}
&\yV =
  \ehmp\yvt
  +\frac{\Ix\tm}{\Cm} \bigl(1-\ehmp\bigr) 
 + \frac{\Is\tm}{\Cm}\ts\ehsm\frac{\ehmp-\ehsp}{\tm-\ts} 
.
\end{align}

The set of critical points of the map $\ybt$ is in this case a 1-dimensional
curve, given in parametric form by
\begin{equation}
  \label{eq:2D_surface_parametric}
  \ybo\colon t \mapsto
  \biggl(
  \etsp \Bigl(\frac{\yvt\Cm}{\tm}-\Ix\Bigr),\quad
\frac{\tm \etmp -\ts \etsp}{\tm-\ts}\frac{\tm}{\Cm}\Bigl(\frac{\yvt\Cm}{\tm}-\Ix\Bigr)
+\frac{\tm}{\Cm}\Ix
\biggr);
\end{equation}
the coordinate $\yv$ of the general
form~\eqref{eq:equation_boundary_implicit} do not exist in this case,
because the threshold is 1-dimensional. \Ref{fig:bounding_surface_jacobian} visualizes the set by the green curve.

\subsubsection{Hypervolume boundary in implicit form}
\label{sec:boundary_implicit_example}

The next step in our procedure is to convert the parametric
equation~\eqref{eq:2D_surface_parametric} of the \env\ into an
explicit or implicit equation for the coordinates $(\Is, \yV)$. As
\Ref{sec:hyperv_boundary_implicit} does not provide a general
algorithm, below we illustrate the process using our example model.

By equating the first component of \eqn~\eqref{eq:2D_surface_parametric} to
the coordinate $\Is$ and solving for $t$, we find
\begin{subequations}
  \begin{gather}
    \label{eq:implicit_form_t_out}
    t = -\ts \ln \Bigl[
    \Bigl(\frac{\yvt\Cm}{\tm}-\Ix\Bigr)/I\Bigr],
    \\
    \intertext{subject to the condition for $\Is$}
    \label{eq:condition_current}
    \e^{-\frac{\yh}{\ts}} \le \frac{1}{\Is} \Bigl(\frac{\yvt\Cm}{\tm}-\Ix\Bigr) \le 1,
  \end{gather}
\end{subequations}
required for $0 \le t \le \yh$ and a real logarithm.

Substituting \eqn~\eqref{eq:implicit_form_t_out} into the $\yV$ coordinate
of \eqn~\eqref{eq:2D_surface_parametric} we find
\begin{equation}
  \label{eq:explicit_V}
  \yV=
\frac{\tm \Ix}{\Cm}
+ 
\frac{\tm \Is}{\Cm}\,
  \frac{\tm
   \bigl[
   \bigl(\frac{\yvt\Cm}{\tm}-\Ix\bigr)/\Is
   \bigr]^{1-\frac{\ts}{\tm}}
    -\ts
  }{\tm-\ts}
\end{equation}
subject to the condition~\eqref{eq:condition_current}. Let us analyse this
equation, in view of its extension to an inequality of the form
$\yS(\Is,\yV)<0$ as required in \Ref{sec:hyperv_boundary_implicit}.
First, we observe that
\begin{equation}
  \label{eq:condition_fraction_exponentials}
  \frac{\tm \etmp -\ts \etsp}{\tm-\ts} \le 1
  \quad\text{for}\quad 0 \le t \le \yh.
\end{equation}
The inequality can be proven by studying the derivative of the
fraction with respect to $t$. The derivative is always negative in
the range above and the only maximum of the fraction is the value unity
assumed at $t=0$.

Inspection of  equation~\eqref{eq:explicit_V} and of its parametric
form~\eqref{eq:2D_surface_parametric} shows that we must consider three
cases: $\Ix \lesseqgtr \yvt\Cm/\tm$, \ie\ whether the external current is
smaller or larger than the \emph{rheobase current}
\begin{equation}
  \ir=\yvt\Cm/\tm;
  \label{eq:rheobase_current_def}
\end{equation}
this is the current necessary to reach threshold in an infinite time
starting from any state with $\Is\le 0$.

\smallskip
\begin{itemize}[para]
\item If $\Ix < \ir$, then $\Is$ is restricted to
  \begin{equation}
    \label{eq:restriction_I_positive}
   0 < \ir-\Ix \le \Is \le
    \e^{\frac{\yh}{\ts}} (\ir-\Ix).
  \end{equation}
  In this case, using the
  inequality~\eqref{eq:condition_fraction_exponentials}, the $\yV$
  component of the \env~\eqref{eq:2D_surface_parametric} is
  always smaller than the threshold:
  \begin{equation}
    \label{eq:V_positive_case}
    \yV \equiv
    \frac{\tm \Ix}{\Cm}
    + 
    \frac{\tm \Is}{\Cm}\,
    \frac{\tm
      \bigl[
      (\ir-\Ix)/\Is
      \bigr]^{1-\frac{\ts}{\tm}}
      -\ts
    }{\tm-\ts}
    \le \yvt
    \quad
    \text{when}\quad \Ix < \ir.
  \end{equation}

  \item If $\Ix > \ir$, then $\Is$ must be negative and
  restricted to
  \begin{equation}
    \label{eq:restriction_I_negative}
    \e^{\frac{\yh}{\ts}} (\ir-\Ix)
    \le \Is \le
        \ir-\Ix < 0.
  \end{equation}
  In this case, using the
  inequality~\eqref{eq:condition_fraction_exponentials}, the $\yV$
  component of the \env~\eqref{eq:2D_surface_parametric} is
  always larger than the threshold:
  \begin{equation}
    \label{eq:V_negative_case}
    \yV \equiv
    \frac{\tm \Ix}{\Cm}
    + 
    \frac{\tm \Is}{\Cm}\,
    \frac{\tm
      \bigl[
      (\ir-\Ix)/\Is
      \bigr]^{1-\frac{\ts}{\tm}}
      -\ts
    }{\tm-\ts}
    \ge \yvt
    \quad
    \text{when}\quad \Ix > \ir.
  \end{equation}

\item If $\Ix =\ir$, the \env\ degenerates to a point: $\ybo(t)=(0,\yvt)$,
  which is the limit point reached in infinite time from any initial point
  in the \nospike\ region; this is the geometric interpretation of the
  equality of external and rheobase currents.
\end{itemize}

\medskip The function representing the \env,
\eqn~\eqref{eq:curved_surf_implicit_equality}
of \Ref{sec:hyperv_boundary_implicit}, has in this case the explicit form
\begin{multline}
  \label{eq:implicit_curve_V-I}
  \yS(\Is,\yV) \defd
  \yV - \frac{\tm \Ix}{\Cm}
- 
\frac{\tm \Is}{\Cm}\,
\frac{\tm
  \bigl[
      (\ir-\Ix)/\Is
   \bigr]^{1-\frac{\ts}{\tm}}
    -\ts
  }{\tm-\ts},\\
  \text{with }(\Is,\yV) \in \yD \defd
  \clcl*{\ir-\Ix ,
\e^{\frac{\yh}{\ts}} (\ir-\Ix)}\times\RR.
\end{multline}
If we calculate its differential, as discussed in
\Ref{sec:hyperv_boundary_implicit}, we find that the latter is a
positive multiple of the differential of the backpropagated threshold
for each $t$. This is true in each of the three cases
above. Consequently,
the inequality $\yS(\Is,\yV) < 0$, $ (\Is,\yV)
\in \yD$ always includes the \nospike\ region.

In summary, we arrive at an analytic, implicit equation for the curved
boundary~\eqref{eq:implicit_curve_V-I}.  The expression is a
transcendental function in $\Is$, owing to the generally irrational
exponent $1-\ts/\tm$, but \emph{it is used in an inequality}, hence we
do not need to find its roots.
This is in contrast to the original transcendental equation for the
threshold-crossing condition~\eqref{eq:threshold_cross_time_example},
which requires a bisection algorithm to find its solution.

\subsubsection{Boundary intersections and final system}
\label{sec:boundary_intersections_example}

We can now assemble the system of inequalities defining the
\nospike\ region consisting of the
boundaries~\eqref{eq:boundary_t0_example}
and~\eqref{eq:boundary_th_example}, and the
\env~\eqref{eq:implicit_curve_V-I}.

\begin{figure}[!hbt]
  \centering
\includegraphics[width=0.8\linewidth]{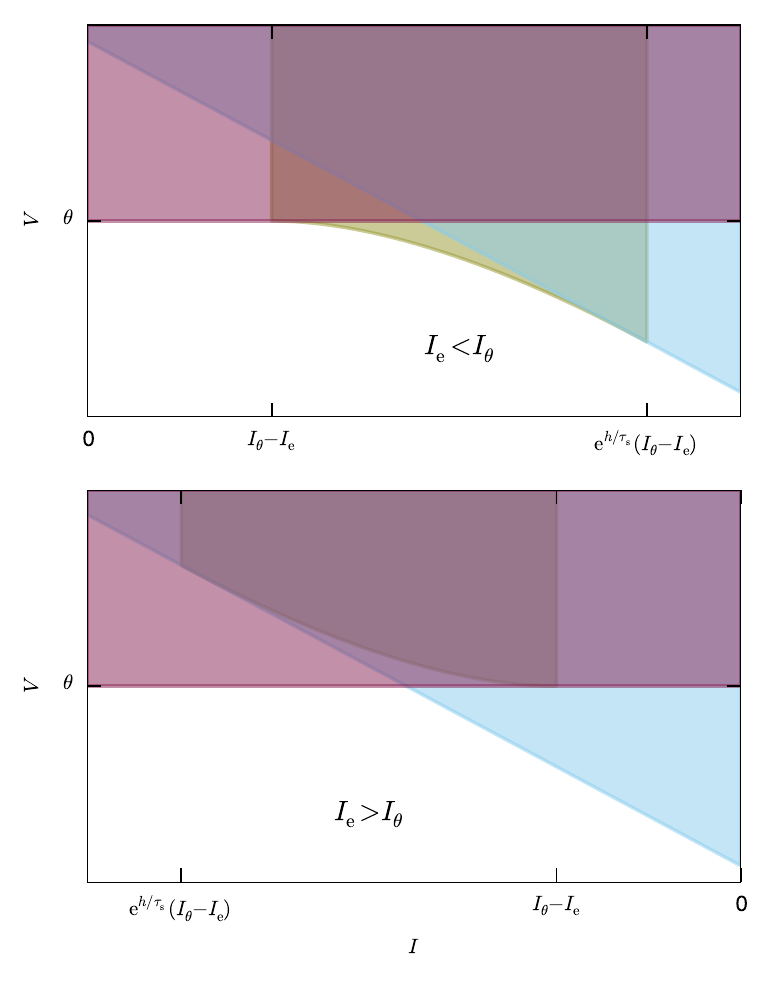}
\caption{System of inequalities, \eqn~\eqref{eq:main_inequalities_2D},
  determining the \nospike\ region. The colored areas represent the
  complementary inequalities of the system~\eqref{eq:main_inequalities_2D},
  so the solution of that system is the white area. The red region
  delimited by the horizontal line corresponds to the first equation of the
  system, the blue region delimited by the inclined line to the second, and
  the yellow region delimited by the curved line to the third.
  \textbf{Upper panel:} case $\Ix < \ir$, all inequalities necessary.
  \textbf{Lower panel:} case $\Ix > \ir$, the third inequality is
  redundant.}
  \label{fig:final_system_example}
\end{figure}

With some rearrangements, simplifications, and the introduction of two new
functions $f$, $\yC$, the condition reads:
\begin{shaded}
  \begin{subequations}
      \label{eq:main_inequalities_2D}
      \begin{align}
           \intertext{$\yso= (\Is,\yV) \in \text{\nospike\ region} \Liff{}$}
        &\yV < \yvt
          \label{eq:1st_inequality_horizontal}
        \\
        \intertext{and}
      &\yV < f_{\yh,\Ix}(\Is) \defd
        \ehmp \yvt
        +\frac{\tm\Ix}{\Cm}\bigl(1-\ehmp\bigr) 
 +\ts\ehsm \frac{\tm\Is}{\Cm}\frac{\ehmp-\ehsp}{\tm-\ts}
          \label{eq:2nd_inequality_inclined}
\\
        \intertext{and}
 &
\yV < \yC_{\Ix}(\Is) \defd \frac{\tm \Ix}{\Cm}
+
\frac{\tm \Is}{\Cm}\,
  \frac{\tm
    \bigl[
      (\ir-\Ix)/\Is
   \bigr]^{1-\frac{\ts}{\tm}}
    -\ts
   }{\tm-\ts}\quad
   \text{ if }
   \Is \in
   \clcl*{\ir-\Ix ,
\e^{\frac{\yh}{\ts}} (\ir-\Ix)}
          \label{eq:3rd_inequality_curved}
    \end{align}
  \end{subequations}
\end{shaded}
\Ref{fig:final_system_example} illustrates the system for the two
cases $\Ix < \ir$ and $\Ix > \ir$. In the former case all three
inequalities are necessary; in the latter, as well as for $\Ix = \ir$,
the last inequality is automatically enforced by the first because its
right-hand side is larger than the threshold $\yvt$ (see
\eqn~\eqref{eq:V_negative_case}).

\subsection{The region of missed spikes}
\label{sec:region_missed_spikes_2D}

In \Ref{fig:bounding_surface_jacobian}, lower panel, the trajectories of
several states during a timestep $\yh$ and ending on the threshold line are
represented by yellow curves. In that figure we can identify a region were
such trajectories self-intersect: it is bounded by a segment of the thick
light blue line, a segment of the thick dark blue line, and a portion of
the thick green curved line. Trajectories with initial states in this
region must therefore cross the threshold twice during the interval
$\opcl{\yto,\yto+\yh}$. As explained in \Ref{sec:region_missed_spikes},
their threshold-crossing is not detected by the sole condition
$\yV[\ys(t+\yh)] \ge \yvt$. All states in this region thus generate
spikes that are missed by the standard test \eqref{eq:test_above_threshold}.

This region is crucial for the comparison of the performances of schemes
implementing the present \sda\ and schemes relying on the standard
test~\eqref{eq:test_above_threshold}. This comparison is quantitatively
made in \Ref{sec:comparison_old_model}. 

\clearpage
\subsection{Optimization, time performance, and accuracy}

\subsubsection{Numerical implementation and optimization}
\label{sec:optimization_sequence}

In the last section we arrived at the system of three
inequalities \eqref{eq:main_inequalities_2D} that determines whether
or not the current state $(\yV, \Is)$ will cross the threshold within
the timestep $\yh$. The state will cross the threshold if the system
is not satisfied, and it will not cross the threshold if the system is
satisfied. This system of inequalities constitutes the \sda\ in the
present model. The system requires the current timestep $\yh$, state
$(\yV, \Is)$, and external electric current $\Ix$ as inputs (cf.\
\Ref{sec:solution}). The timestep $\yh$ is the minimum between the
global timestep and the time interval up to the next input coming from
other neurons, hence it can differ every time the test is called. The
external electric current $\Ix$ may also vary, stepwise, during the
simulation, hence it may also be distinct at each call of the test.

If the system of inequalities is satisfied, thus predicting the absence of a spike within
a timestep $\yh$, then the state of the neuron is evolved by applying the
propagator~\eqref{eq:evolution_initial_state} with
\eqref{eq:2D-case_correspondence}--\eqref{eq:exponential_A}, leading to a
new state, and the procedure starts again. If the system is not satisfied,
thus predicting the occurrence of a spike within $\yh$, it is then necessary to
compute the time $\ytte$ at which the threshold is crossed. This
calculation, explained in Appendix~\ref{sec:membrane_potential_dynamics},
is made by interpolation between the current state and time, and the state
and time $\ytm$ at which the membrane potential would reach its maximum if
allowed to increase above threshold.
Once $\ytte$ is calculated, a spike is emitted, communicated to the
postsynaptic neurons, and the membrane potential is reset to
$V_\text{reset}$ for a refractory period. 
When this refractory period is over the procedure starts again.

In the evolution loop just described, the membrane potential is reset to a
value below threshold as soon as it crosses the latter. Thus no initial
state can have $\yV \ge \yvt$. This means that the first inequality
\eqref{eq:1st_inequality_horizontal} in the system is always satisfied and
can be dropped. Only inequalities \eqref{eq:2nd_inequality_inclined}
and \eqref{eq:3rd_inequality_curved} have to be assessed,
leading to the
reduced system
\begin{equation}
  \label{eq:reduced_inequalities_2D}
  \left\{
    \begin{aligned}
      \yV &< f_{\yh,\Ix}(\Is),\\
      \yV &< \yC_{\Ix}(\Is).
    \end{aligned}
\right.
\end{equation}
We first discuss the case $\Ix < \ir \equiv \yvt\Cm/\tm$.

\begin{figure}[!tp]
  \centering
    \includegraphics[width=\textwidth]{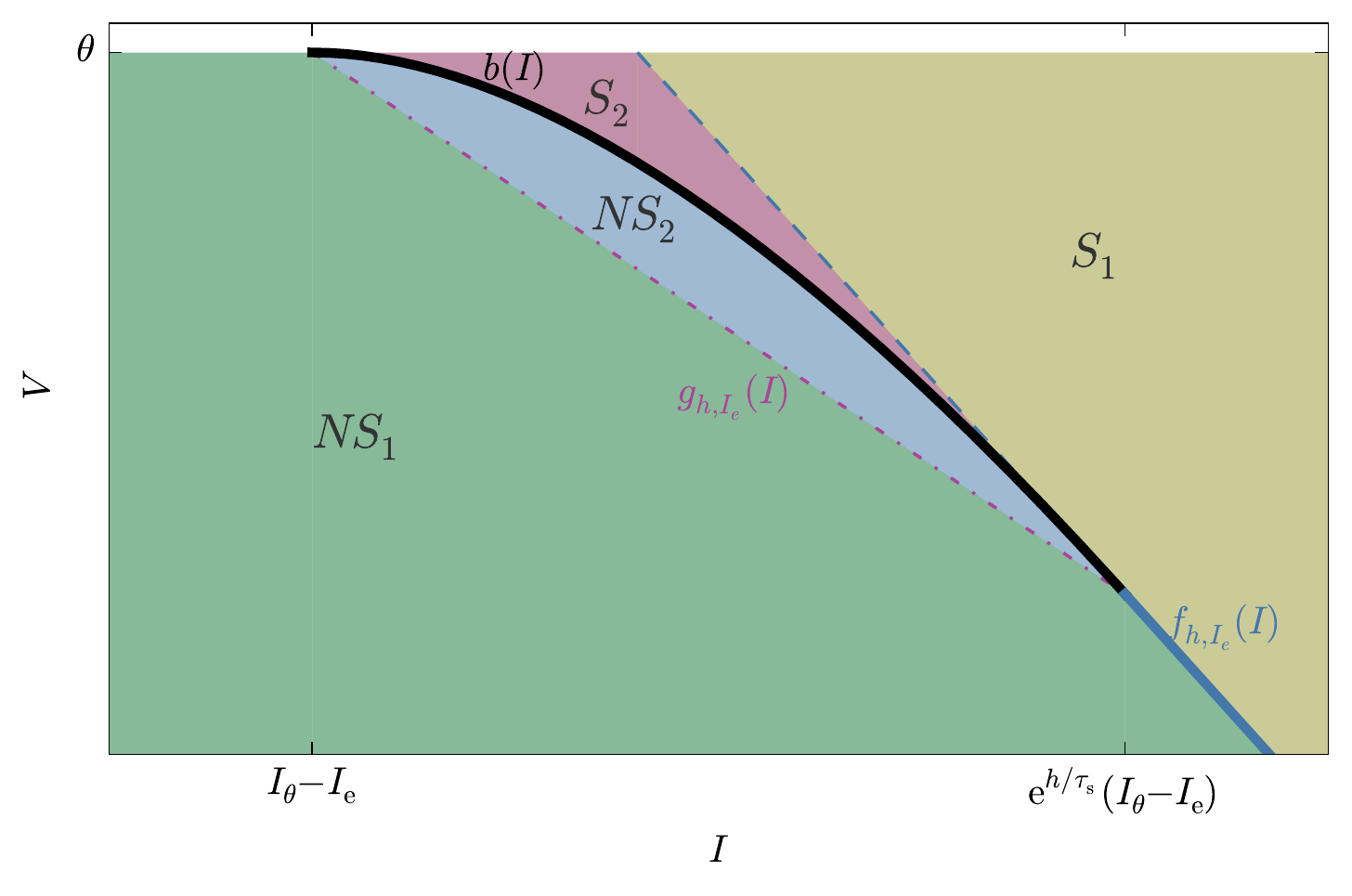}
    \caption{ State-space subregions formed by the intersections of the
      reduced inequalities~\eqref{eq:reduced_inequalities_2D}:
      $\yV < f_{\yh,\Ix}(\Is)$ (straight blue line, partly dashed partly
      continuous) and $\yV < \yC_{\yh}(\Is)$ (black curve), and by the
      auxiliary inequality~\eqref{eq:auxiliary_inequality}:
      $\yV < g_{\yh,\Ix}(\Is)$ (dot-dashed straight purple line). The \spike\ region is
      $\Sp_1 \cup \Sp_2$, the \nospike\ region is $\NSp_1 \cup \NSp_2$. The
      subregion $\Sp_2$ contains all states that emit spikes undetected by
      the standard test~\eqref{eq:test_above_threshold}; they are detected
      by the \sda.}
\label{fig:spiking-region}
\end{figure}

The geometric meaning of the inequalities above is illustrated in
\Ref{fig:spiking-region}. The figure shows four regions: $\NSp_1$,
$\NSp_2$, $\Sp_1$, $\Sp_2$. The \nospike\ region is the union of $\NSp_1$
and $\NSp_2$, the \spike\ region the union of $\Sp_1$ and $\Sp_2$. In the
figure they are separated by a thick line, partly curved and black, partly
straight and blue. Subregion $\Sp_2$ is particularly important: it is the
region of missed spikes discussed in \Ref{sec:region_missed_spikes_2D},
corresponding to the self-intersection region of
\Ref{fig:bounding_surface_jacobian}, lower panel. It contains those states
that lead to spikes missed by schemes that rely on the standard
test~\eqref{eq:test_above_threshold}.

Region $\Sp_1$ is separated from $\Sp_2$ by a dashed blue line, and from
$\NSp_1$ by a continuous blue line, the continuation of the dashed one.
This partly dashed, partly continuous blue line corresponds to the equation
$\yV = f_{\yh,\Ix}(\Is)$. Hence if the inequality $\yV < f_{\yh,\Ix}(\Is)$
is \emph{not} satisfied then the initial state is in region $\Sp_2$ or on its
blue boundary, and there will be a spike. If the inequality is satisfied
the state could be in $\Sp_2$ -- spike -- or $\NSp_1 \cup \NSp_2$ -- no
spike; an undetermined case. This inequality requires modest computational
costs because it is linear in $\Is$ and $\Ix$ and involves exponentials of
$\yh$. Regions $\Sp_2$ and $\NSp_2$ are separated by a black curve: this is the
\env, corresponding to the equation $\yV = \yC_{\Ix}(\Is)$ for
$\Is \in \clcl*{\ir-\Ix , \e^{\frac{\yh}{\ts}} (\ir-\Ix)}$. Hence if the
inequality $\yV < \yC_{\Ix}(\Is)$ is \emph{not} satisfied the initial state
is in $\Sp_2$ or on its boundary, and there will be a spike. If the
inequality is satisfied the initial state is either in $\NSp_2$, or in
$\NSp_1$ with $\ir-\Ix < \Is < \e^{\frac{\yh}{\ts}} (\ir-\Ix)$, and no
spike will occur.

The computationally most expensive inequality is $\yV < \yC_{\Ix}(\Is)$
because it involves irrational powers of $\Is$ and $\Ix$. It is advisable
to avoid its direct computation as often as possible by pre-testing a
linear inequality. In \Ref{sec:convexity_approx} we discussed how such a
pre-test is indeed possible thanks to the convexity of the \nospike\
region. There, we argued that the curved \env\ can be approximated by
triangular hypersurfaces, which simply reduce to one straight segment in
the present two-dimensional case: this is the dot-dashed red line in
\Ref{fig:spiking-region}, separating $\NSp_1$ and $\NSp_2$. This line has
equation $\yV = g_{\yh,\Ix}(\Is)$ with
\begin{equation} \label{eq:sec-line}
 g_{\yh,\Ix}(\Is) \defd
  \yvt+
 \frac{\ts  \e^{\frac{\yh}{\ts}}}{\tm-\ts}  \frac{\tm}{\Cm}\Is
  +\e^{\frac{\yh}{\tm}}\,\frac{\tm}{\Cm} (\ir -\Ix),
\end{equation}
and the corresponding inequality
\begin{equation}
  \yV < g_{\yh,\Ix}(\Is)\label{eq:auxiliary_inequality}
\end{equation}
has the same computational costs as $\yV < f_{\yh,\Ix}(\Is)$.

If the auxiliary inequality $\yV < g_{\yh,\Ix}(\Is)$ is satisfied, the
initial state is in $\NSp_1$ and $\yV < \yC_{\Ix}(\Is)$ is also satisfied.
It is therefore convenient to test the auxiliary inequality before the
computationally costly one, which can be discarded if the test is
positive. \Ref{fig:spiking-region} suggests that this test might be
positive for the majority of initial states because region $\NSp_1$ is
much wider than $\NSp_2$. This possibility would be very advantageous, but
we now argue that it should be verified by a dynamical analysis.

The system \eqref{eq:reduced_inequalities_2D} can be translated into a
computational algorithm in several different ways, depending on the order of
evaluation of its two inequalities and of the auxiliary
inequality~\eqref{eq:auxiliary_inequality}. In simplified terms, such an
algorithm consists in a sequence of tests -- variously implemented as
\texttt{if}, \texttt{and}, \texttt{or} constructs -- for finding the
initial state in space-time regions $R_1$, $R_2$, and so on. The order of
these tests is important. The average time cost of the algorithm in a long
simulation is given by $\sum_i p_i c_i$, where $p_i$ is the frequency with
which states are found in region $R_i$, which we call \enquote{occupation
  frequency}, and $c_i$ is the cumulative time cost of the test for region
$R_i$. This time cost $c_i$ is cumulative in the sense that all tests up to
the $(i-1)$th must have been performed, with \texttt{false} outcomes, to
arrive at the test for $R_i$. The efficiency of an algorithm therefore
depends on the mathematical form of the inequalities defining a region and
on the occupation frequencies of the regions, determined by the dynamics.
These two factors can be extrapolated by a theoretical analysis, or more
practically measured by running long test simulations with typical network
setups corresponding to the cases one is interested in.

We now try to determine the most efficient algorithm for the present case.
Region $\Sp_1$ is the least costly, because bounded by one line and
therefore involving one inequality linear in $\Is$; then region $\NSp_1$,
bounded by two lines involving two linear inequalities; and finally regions
$\NSp_2$ and $\Sp_2$, bounded by the curve that involves rational
exponentiation. For this example model, we tried different orderings
but show here only two possible extreme cases to illustrate that there is no
significant difference in the computational cost.
\\[-2\jot]
\hspace*{\fill}
\begin{minipage}[t]{0.45\textwidth}
  \begin{algorithm}[H]
  \caption{\protect\centering\label{alg:model-1}}
\begin{algorithmic}
\State \textbf{bool} is\_spike($\yh$):
\newline pre-compute $g_{\yh,\Ix}(\Is)$ 
\If {$V \le  f_{\yh,\Ix}(\Is)$ \textbf{and} $V< g_{\yh,\Ix}(\Is) $}
    \State return \textbf{false}  
\ElsIf{$V \ge g_{\yh,\Ix}(\Is)$}
        \State return \textbf{true} 
\ElsIf{$V \ge \yC_{\Ix}(\Is)$}
  \State return \textbf{true} 
\Else
  \State return \textbf{false} 
\EndIf
\end{algorithmic}
  \end{algorithm}
\end{minipage}
\hfill
\begin{minipage}[t]{0.4\textwidth}
  \begin{algorithm}[H]
  \caption{\protect\centering\label{alg:model-2}}
  \begin{algorithmic}
  \State \textbf{bool} is\_spike($\yh$):
  \If {$V \ge  g_{\yh,\Ix}(\Is)$ \textbf{or} \\\hspace*{\fill}
    [$V \ge f_{\yh,\Ix}(\Is)$ 
   \textbf{and} $V \ge \yC_{\Ix}(\Is)$]}
      \State return \textbf{true}  
  \Else
    \State return \textbf{false} 
  \EndIf
  \end{algorithmic}
  \end{algorithm}
\end{minipage}
\hspace*{\fill}

\bigskip

\alg~\ref{alg:model-1} is based on the assumption that the occupation
frequency of a subregion is proportional to that subregion's relative size.
If we check the two largest first, in the order $\NSp_1$, $\Sp_1$, $\Sp_2$,
$\NSp_2$, we are therefore more likely to exit the test in its first
\texttt{if} branches.
The value of $g_{h,\Ix}$ is used in two \texttt{if} branches,
so it is computed just once and saved before the \texttt{if} sequence in
order to save some computations.

\alg~\ref{alg:model-2} uses a composite \texttt{or}-\texttt{and} condition
rather than several \texttt{if}s. Assuming left-to-right evaluation, the
algorithm corresponds to testing first $\Sp_1$ (left side of \texttt{or}),
then $\Sp_2$ (right side of \texttt{or}), either case leading to a spike.
If neither is true, no further tests are necessary because the state must
necessarily be in the \nospike\ region. The test for $\Sp_2$ is made less
costly on average by using the auxiliary
inequality~\eqref{eq:auxiliary_inequality}. This algorithm uses one test
less overall than the previous one, but it may require one more test on
average, if $\NSp_1$ is the region with highest occupation frequency. The
left-to-right evaluation assumption does not always hold in modern
processors, which build their own statistics to optimize the test order of
logical constructs.

\smallskip

The analysis assumed $\Ix < \ir \equiv \yvt\Cm/\tm$. The reduced system
\eqref{eq:reduced_inequalities_2D} and the \alg~\ref{alg:model-1}
and~\ref{alg:model-2} are, however, also valid in the case that
$\Ix \ge \ir $, corresponding to the lower panel of
\Ref{fig:final_system_example}. In this case subregions $\NSp_2$ and
$\Sp_2$ do not exist below the threshold, and the inequalities
$\yV < \yC_{\Ix}(\Is)$ and $\yV < g_{\yh,\Ix}(\Is)$ are always satisfied
when $\yV < \yvt$; they always evaluate to \texttt{true} in both
algorithms. Both algorithms therefore correctly distinguish spiking from
non-spiking states in this case, although they become inefficient owing to
the additional superfluous evaluations of $\yC_{\Ix}(\Is)$ and
$g_{\yh,\Ix}(\Is)$ for spiking states. We decide not to modify them in the
present work because the case $\Ix \ge \ir$ is unusual in real
applications. More efficient algorithms for this case can be designed by
interested readers following the guidelines just given in this section.

\subsubsection{Occupation frequencies}
\label{sec:occupation_freqs}

We want to measure the occupation frequencies in the typical case of a
neuron embedded in a recurrent network, receiving fluctuating synaptic
input. The setup for this simulation is illustrated in \Ref{fig:network}A
and its formulae explained in Appendix~\ref{sec:noisy_network_setup}. One
neuron is coupled with strengths $\yj$ and $-\yj$ to an excitatory and an
inhibitory Poisson generator and also receives a constant external current.
The Poisson generators each mimic an excitatory and an inhibitory population.
The neuron thus receives a fluctuating input current having average $\mu$
and variance $\yvar$. In this instance the code of the simulation includes
a subroutine that informs us of the current state every time the \sda\ test
is called, without altering the test or the dynamics.

\begin{figure}[!tb]
  \centering
  \includegraphics[width=\textwidth]{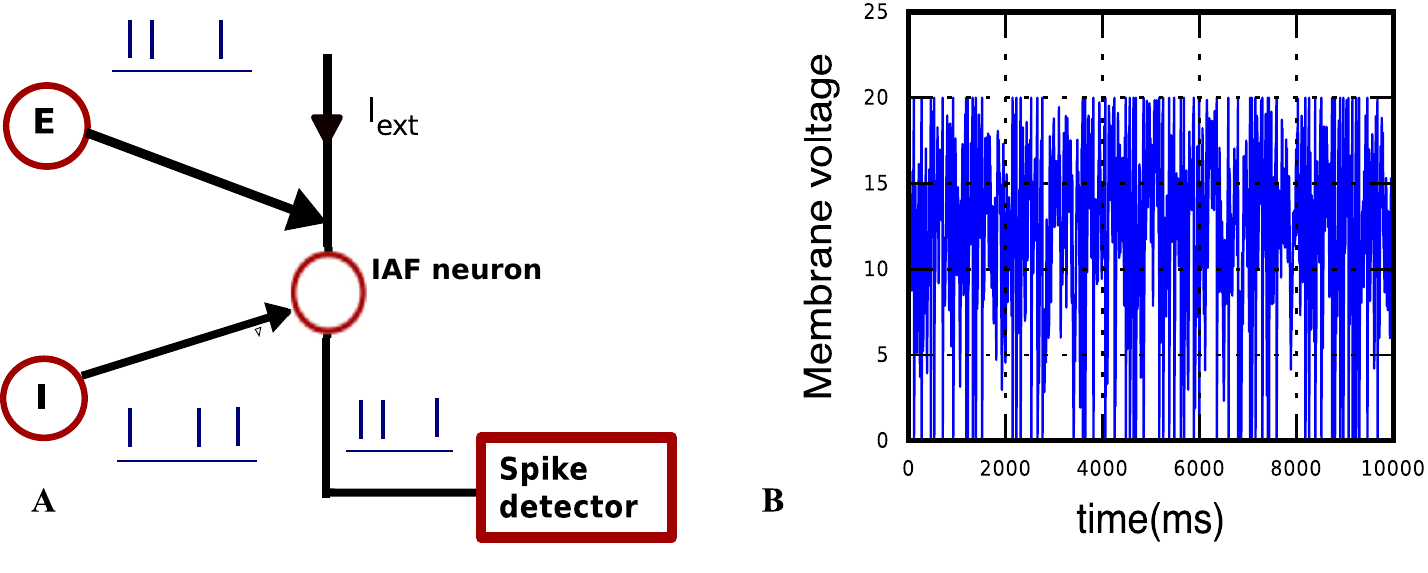}
  \caption{(\textbf{A}) Schematic of the simulation setup used to calculate
    occupation frequencies and to compare the hybrid scheme with \sda\
    \alg~\ref{alg:model-1}, with \sda\ \ref{alg:model-2}, and with the standard
    test. The neuron model (empty red circle), implementing one of the three
    schemes, receives input from an external current and from one
    excitatory (E) and one inhibitory (I) Poisson generator. The total
    input has mean $\mu$ and variance $\yvar$. (\textbf{B}) Sample of
    membrane-potential dynamics for $\mu= 15 \umV$ and
    $\yvar = 25\;\textrm{mV}^2$.}
\label{fig:network}
\end{figure}

\begin{figure}[!tb]
  \centering
  \includegraphics[width=\textwidth]{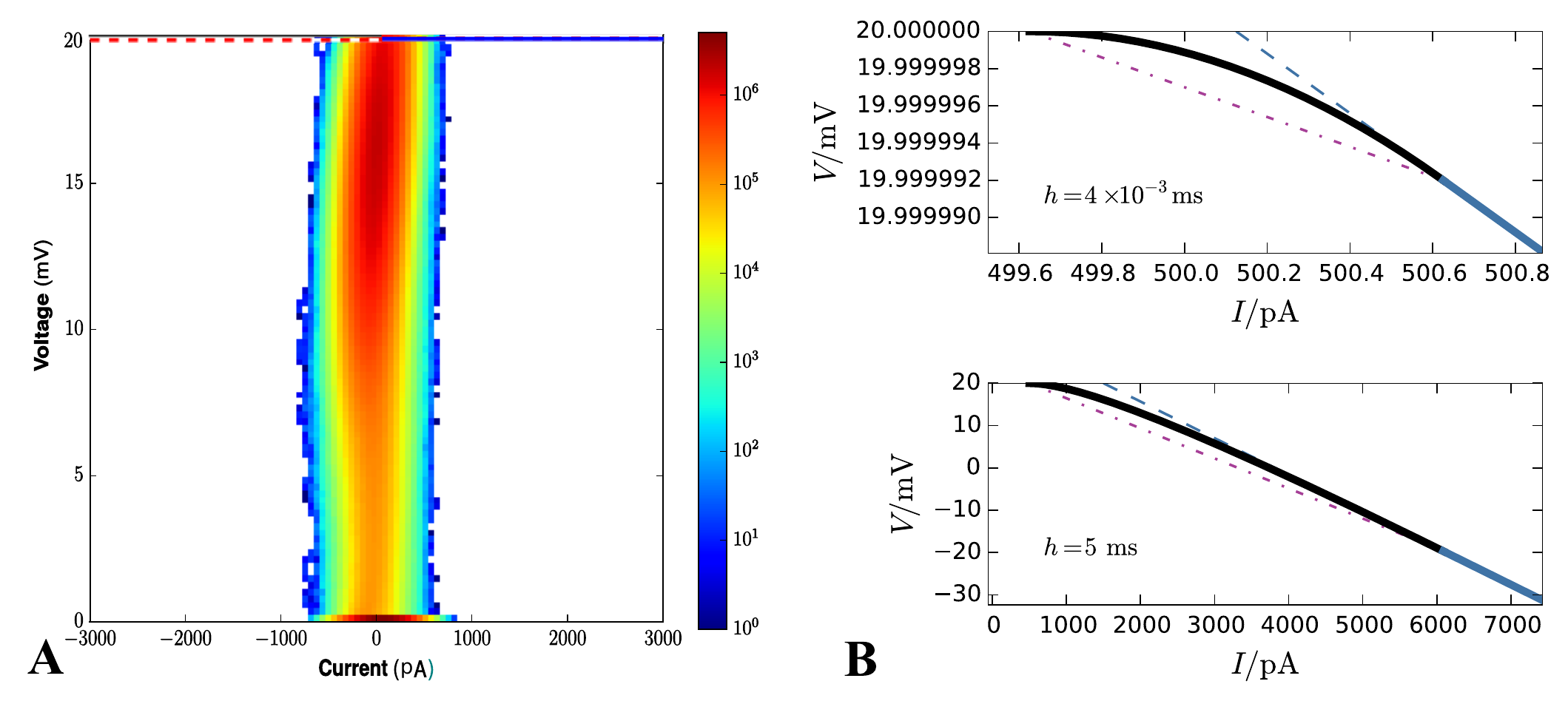}
  \caption{(\textbf{A}) Frequency density of states over state space at
    each call of the threshold-crossing test, for network parameters $\mu=15 \umV$,
      $\yvar = 25\;\textrm{mV}^2$, $\yj = 0.1 \umV$. The colourbar is in units of
    $6\times 10^9\;\textrm{mV}^{-1}\,\textrm{pA}^{-1}$ (obtained from total number of
    events \texttimes\ area element). The dotted purple line
    is the threshold $\theta$. The only visible region in this plot is $\NSp_1$.
    and the horizontal blue line is the boundary between
     \nospike\  and \spike\ regions. The \spike\ region $\Sp_1 \cup \Sp_2$
     and subregion
    $\NSp_2$  are not visible on this scale because of the
    exceedingly small average timestep $\yh = 4 \times10^{-3}\ums$.
To discern them we need to zoom in, as done in upper panel (\textbf{B}): the curved \env\ and the two straight lines
    that separate $\Sp_1$, $\Sp_2$, and $\NSp_2$ extend horizontally and vertically for just
    about $1\;\textrm{pA}$ and  $10^{-5}\umV$. (\textbf{B}) lower panel. In
    contrast, for a much larger
    timestep $h=5\ums$ the three boundaries would have a larger  extension,
    about $5\,500\;\textrm{pA}$ and $20\umV$, and be discernible in plot (A).
  }
\label{fig:histogram}
\end{figure}

\Ref{fig:histogram}A gives a visual idea of the occupation frequencies for
$\mu=15\umV$, $\yvar=25\;\textrm{mV}^2$, and $\yj= 0.1\umV$ (all expressed
in volts through multiplication by a resistance of
$\tm/\Cm = 40\;\textrm{M\textohm}$), which correspond to the case
$\Ix < \ir$. These values correspond to a composite average input of
$250\,000\;\mathrm{spikes/s}$, and a total average input current of
$400\;\textrm{pA}$. This presynaptic input makes the neuron fire at an
average rate of $7\;\textrm{spikes/s}$. A sample of its membrane dynamics
is shown in \Ref{fig:network}B. Subregion $\NSp_1$ has the overwhelmingly
largest occupation frequency. The other three subregions have actually very
small areas, as clear from the axis ranges of \Ref{fig:histogram}B, owing
to the very small value of the average timestep,
$\yh = 4\times 10^{-3}\ums$, given by the inverse of the input rate in this
hybrid scheme.

\begin{figure}[hbpt]
  \includegraphics[width =\linewidth]{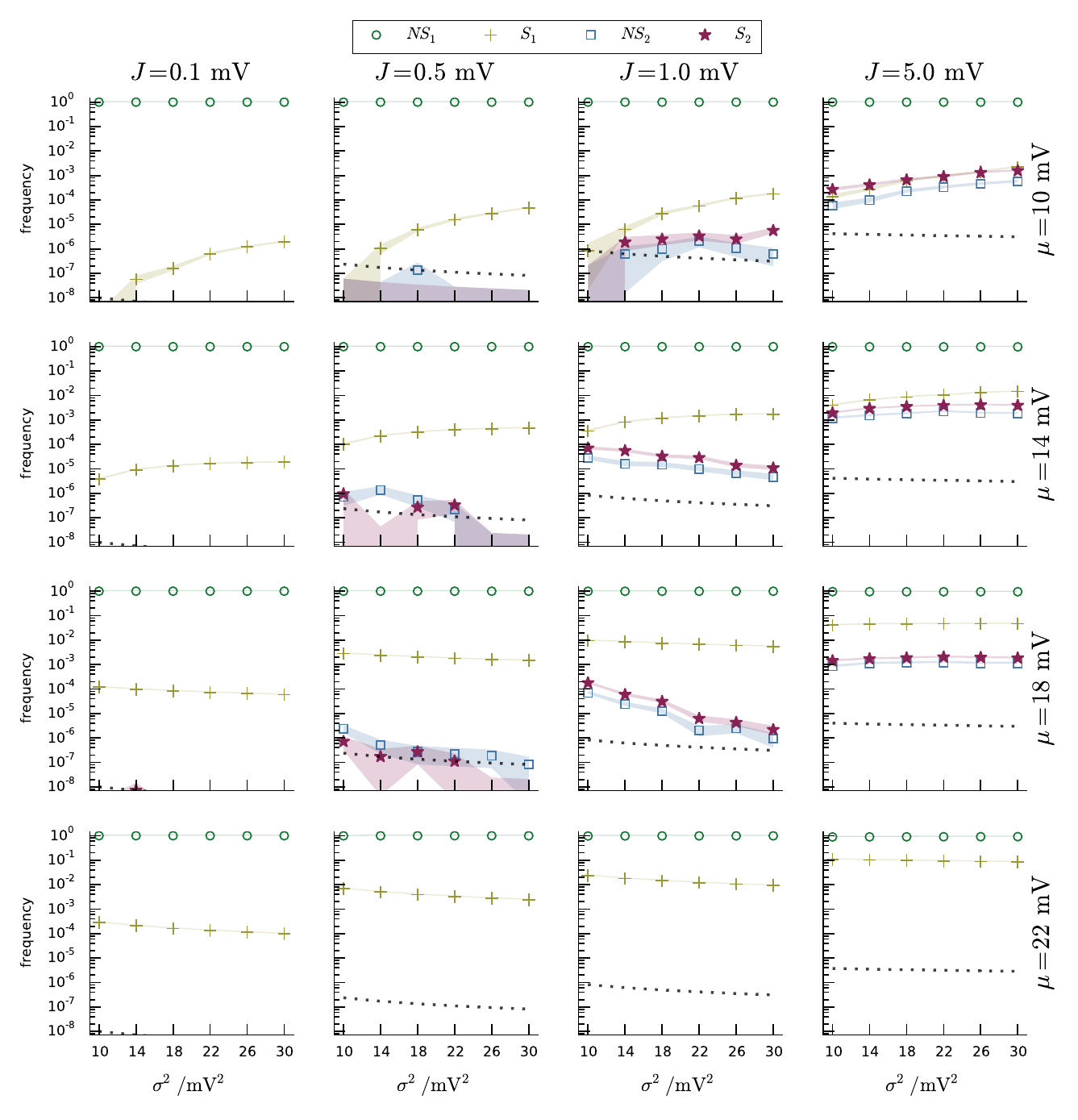}
  \caption{Occupation frequencies of the four subregions $\NSp_1$ (green
    circles), $\Sp_1$
    (yellow crosses), $\NSp_2$ (blue squares), $\Sp_2$ (red
    stars) of \Ref{fig:spiking-region}, for various sets $(\mu,\yvar,\yj)$ of input-current
    mean and variance, and synaptic strength. Each panel shows the frequencies \vs\
    current variance for fixed current mean and synaptic strength. The columns have the same $\yj$, ranging
    from $0.1\umV$ (leftmost) to $5\umV$ (rightmost). The rows have the
    same $\mu$, ranging from $10\umV$ (top) to $22\umV$ (bottom).  The
    frequencies were measured from $N$ samples, depending on
    $(\mu,\yvar, \yj)$. The dotted lines in each plot show the inverse
    number of samples $1/N$ for that network regime. The thickness of the
    segments connecting the data points equals one standard deviation. The
    shaded regions show where the limiting frequencies ($N\to \infty$) are
    expected to lie with $87\,\%$ probability, using a Johnson-Dirichlet
    model with parameter $k=0.05$ determined by posterior maximization
    \citep[\sect~3.2.5
    ]{johnson1932c,zabell1982,good1966,bernardoetal1994_r2000}. 
The frequencies of region $\Sp_2$ (red stars) are
    particularly important: they are the frequencies of spike-misses of the standard test~\eqref{eq:test_above_threshold}. 
  }
  \label{fig:region_frequencies}
\end{figure}

A more precise comparison of the occupation frequencies of the four
regions $NS_1$, $NS_2$, $S_1$, $S_2$ is shown in
\Ref{fig:region_frequencies} for several combinations of three network
parameters, producing different dynamic regimes. The parameters are
the average $\mu$, the variance $\yvar$ of the input current, and the
presynaptic coupling strength $\yj$ (all expressed in volts through
multiplication by a resistance of $\tm/\Cm =
40\;\textrm{M\textohm}$).
The values of the parameters $(\mu, \yvar, \yj)$ include typical
realistic cases as well as some extreme cases, like unusually high
coupling strengths. Each panel of \Ref{fig:region_frequencies} shows
the occupation frequencies for a set of dynamical regimes with
constant $\yj,\ymu$ and several $\yvar$. The panels in the last row
correspond to the case
$\Ix \equiv \mu\Cm/\tm \ge \ir \equiv \yvt\Cm/\tm$, or $\mu \ge \yvt$,
in which subregions $\NSp_2$ and $\Sp_2$ do not exist below threshold.

It is important to remember that the boundary and size of the regions of
\Ref{fig:spiking-region} vary with the timestep $\yh$, which is a parameter
of the simulation scheme, not of the dynamics per se. In an event-driven or
hybrid scheme, this step varies inversely with the event input rate, which
for Poisson input generators is proportional to $\yvar/\yj^2$. As a
consequence, the frequencies displayed in \Ref{fig:region_frequencies} are
not determined by the Liouville distribution of the dynamics
\eqref{eq:dynamics_2D_model} alone, but also by the details of the
numerical-implementation scheme. The dependence of the boundaries on $\yh$
is illustrated in \Ref{fig:histogram}B. As $\yh$ decreases, the line
$\yV = f_{\yh,\Ix}(\Is)$ and the auxiliary line $\yV = g_{\yh,\Ix}(\Is)$
gets closer to the threshold, and the subregions $\Sp_1$,
$\Sp_2$, $\NSp_2$ disappear. This is plausible since in the limit of $\yh=0$ we are not
evolving the initial state at all. As $\yh$ increases the point of tangency
between the \env\ $\yV = \yC_{\Ix}(\Is)$ and the line $\yV = f_{\yh,\Ix}(\Is)$
moves to increasingly lower voltages and higher currents; subregion $\Sp_2$
takes over $\Sp_1$ and subregion $\NSp_1$ becomes wider. For typical
timestep values of several milliseconds, though, subregions $\Sp_1$,
$\Sp_2$, $\NSp_2$ are still very small.

A rough estimate of the dependence of the areas of the bounded regions
$\Sp_2$ and $\NSp_2$ on the parameters $(\mu, \yvar,\yj)$, for $\mu < \yvt$,
can be obtained by looking at \Ref{fig:region_frequencies} and considering
that these areas together form a triangle with vertices
\begin{equation}
  \label{eq:vertices_triangle}
  (\ir-\Ix, \yvt),\quad
  \bigl(\e^{\yh/\ts}\,(\ir-\Ix), \yC_{\Ix}[\e^{\yh/\ts}\,(\ir-\Ix)]\bigr),\quad
  (I_f, \yvt),
  \qquad\text{with $I_f$ such that $f_{\yh,\Ix}(I_f)=\yvt$.}
\end{equation}
This triangle has base $\abs{I_f-(\ir-\Ix)}$ and height
$\abs{\yvt-\yC_{\Ix}[\e^{\yh/\ts}\,(\ir-\Ix)]}$. Expressing $\yh$ and $\Ix$ in
terms of $\mu$ and $\yvar$ using
\eqns~\eqref{eq:mu_sigma_J_r_dependencies}, where $\yh$ is inversely
proportional to the input rate $r_{\text{I}}+r_{\text{E}}$, we find
\begin{equation}
  \label{eq:area_regions_rough}
  \text{areas of $\Sp_2$ and $\NSp_2$}
  \propto
  \frac{\Cm}{\tm}\,
  (\yvt-\mu)^2\,
  \biggl[1-
\frac{\tm\e^{\yh/\tm} - \ts\e^{\yh/\ts}}{\tm-\ts}
\biggr]
\biggl[
\frac{\e^{\yh/\ts}\,(\tm-\ts)\,(1-\e^{\yh/\tm})}{\ts\,(\e^{\yh/\tm}-\e^{\yh/\ts})}
-1
\biggr]
\quad\text{with $\yh=\frac{\tm\yj^2}{\yvar}$.}
\end{equation}
When $\ts\lesssim\tm$ and $\yj^2\lesssim\yvar$ a Taylor expansion in
$\yj^2/\yvar$ to fourth order gives a good approximation, with a relative
error below $10\,\%$:
\begin{equation}
  \label{eq:area_regions_rough_expansion}
\text{areas of $\Sp_2$ and $\NSp_2$}
\propto \frac{\Cm}{\tm}\, (\yvt-\mu)^2\,
\Biggl[
\frac{\tm^2}{4\ts^2}\biggl( \frac{\yj^2}{\yvar}\biggr)^3
+\frac{\tm^2\,(\tm+\ts)}{8{\ts}^3}\biggl( \frac{\yj^2}{\yvar}\biggr)^4
+\Ord\biggl(\frac{\yj^2}{\yvar}\biggr)^5
\Biggr].
\end{equation}
This approximate formula shows that subregions $\NSp_2$ and $\Sp_2$ grow
with the square of the input mean $\mu$ and with the third or fourth power
of the ratio $\yj^2/\yvar$. Recall that these regions do not exist for
$\mu\ge \yvt$. The occupation frequencies do not depend on the areas alone,
however, but also on the dynamics, as explained in the previous section. We
can identify several other dynamical mechanisms for their dependence on the
parameters $(\mu,\yvar,\yj)$:
\begin{itemize}
\item an increase in mean input $\mu$ leads to more frequent threshold
  crossings, thus frequently bringing the voltage to its reset value,
  underneath subregions $\NSp_2$ and $\Sp_2$. The occupation frequencies of
  these subregions may therefore decrease with $\mu$ even though their
  areas grow with $\mu$;
\item for low mean input $\mu$, an increase in variance $\yvar$ means a
  higher chance of high-$\yV$ regions, and thus an increase in the
  occupation frequencies of $\NSp_2$ and $\Sp_2$, even though their areas
  shrink with $\yvar$;
\item for mean input $\mu$ close to the threshold, an increase in the
  variance $\yvar$ leads to more frequent threshold crossings, and may thus
  increase occupation frequency of $\Sp_2$ with $\yvar$,
  even though its area shrinks with $\yvar$.
\end{itemize}

The occupation frequency of subregion $\NSp_1$ (green circles) dominates
all others, varying from $90\,\%$ to $100\,\%$ depending on the network
parameters. Subregion $\Sp_1$ (yellow crosses) follows in order of
frequency and is the most frequently visited between the two \spike\
subregions. Subregions $\NSp_2$ (blue squares) and $\Sp_2$ (red stars) are
scarcely visited for lower synaptic amplitudes, with frequencies from $0$
to $10^{-6}$; and slightly more often at higher synaptic amplitudes
(frequencies from $0$ to $10^{-2}$).

\subsubsection{Time performance and accuracy of a hybrid scheme based on the \sda}
\label{sec:comparison_old_model}

The average time costs of \alg~\ref{alg:model-1} and \alg~\ref{alg:model-2}
within a hybrid scheme can be assessed by real-time simulation
measurements. The average time cost of a hybrid scheme based on the
standard test \eqref{eq:test_above_threshold} can also be
assessed in the same way for comparison. We therefore compare the two
algorithms of the \sda\ and the sufficiency test in this section.

The basic setup is the same as for the frequency analysis of
\ref{sec:optimization_sequence}, explained in
Appendix~\ref{sec:noisy_network_setup}, with network parameters
$(\mu, \yvar, \yj)$. The only difference is that in the present case the
code does not include the subroutine that informs us of the frequencies,
which would otherwise increase and bias the real-time durations of the
simulations. Three instances of the basic setup are prepared: in the first
the neuron is modelled by a hybrid scheme with \sda\
\alg~\ref{alg:model-1}, in the second the neuron is modelled by a hybrid
scheme with \sda\ \alg~\ref{alg:model-2}, and in the third the neuron is
modelled by a hybrid scheme with the standard threshold-crossing
test~\eqref{eq:test_above_threshold}. The various random-number-generator
seeds of the three instances are exactly the same, so that the three
neurons receive exactly the same input, spike-for-spike; this is essential
for a fair comparison between the three schemes. The three instances are
run for a long time ($2\times 10^6\ums$) and repeated (in parallel) for
several times ($10$), enough to collect reliable statistics. The statistics
are collected for the same sets of $(\mu, \yvar, \yj)$ values as in the
frequency analysis. The total real-time length of a simulation depends, in
all three schemes, on how often the checkpoints and threshold-crossing
tests occurs, and this in turn depends on the presynaptic input frequency,
as already discussed.

\begin{figure}[!bt]
  \centering
  \includegraphics[width=0.66\textwidth]{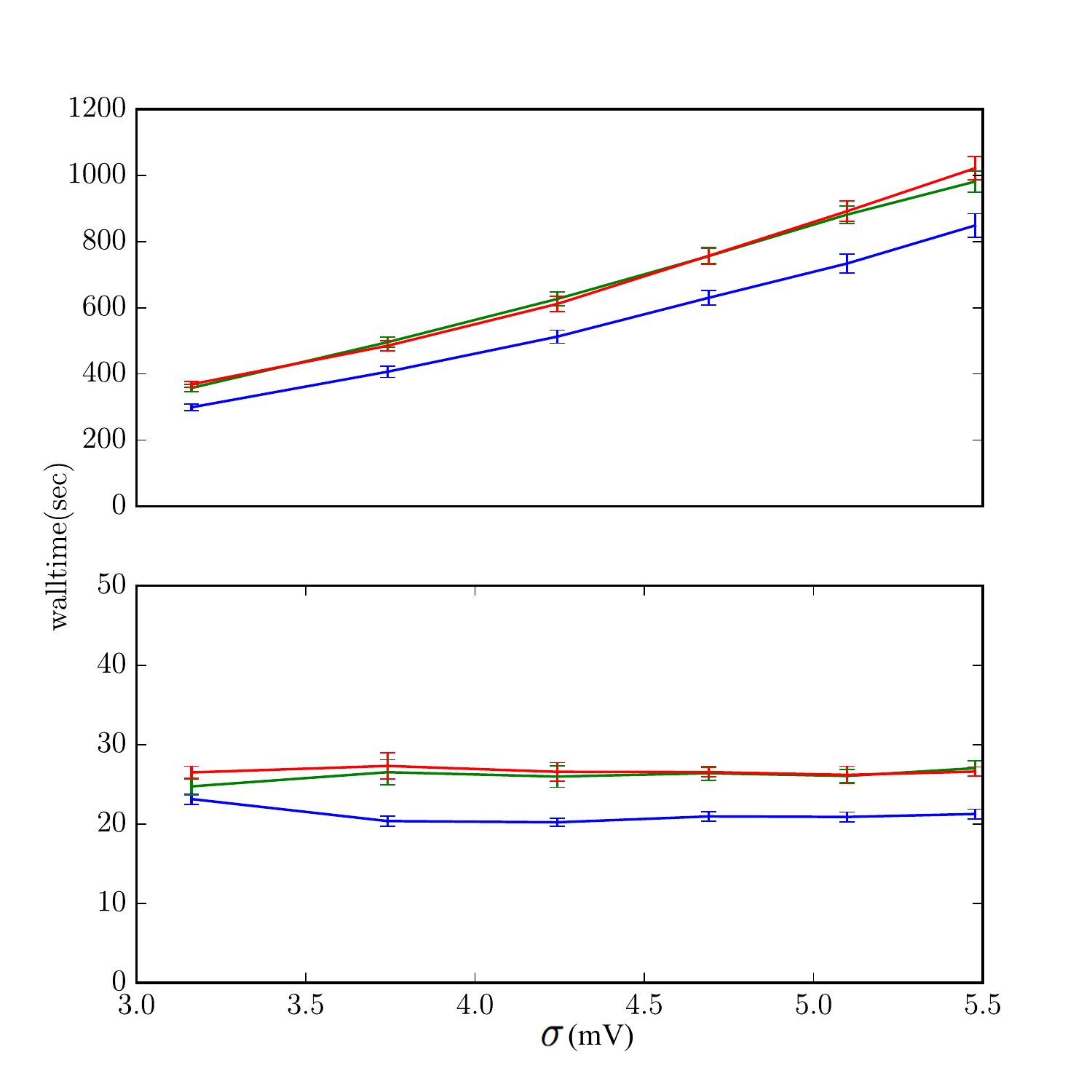}
  \caption{Computational costs of the hybrid scheme with \sda,
    \alg~\ref{alg:model-1} (red), \sda, \alg~\ref{alg:model-2} (green), and
    standard threshold-crossing test~\eqref{eq:test_above_threshold}
    (blue). (\textbf{A}) $\mu=18\umV$, $\yj=0.1\umV$ describes regimes where the \sda\
    has the highest increase in computational cost, around $33\,\%$.
    These are regimes of high activity and frequently incoming spikes.
    (\textbf{B}) $\mu=10\umV$, $\yj=5\umV$ describe regimes where the \sda\
    has the lowest increase in computational cost, around $8\,\%$. These are
    regimes of low activity and infrequently incoming spikes. The data come
    from $10$ simulations of  $2 \times 10^6\ums$ simulation-time each.}
 \label{fig:performance-comparison}
\end{figure}

Rather than showing the results for all sets of parameters
$(\mu, \yvar, \yj)$, which in this case are not very informative, we show
in \Ref{fig:performance-comparison} those with the $(\yj, \mu)$ values that
yield the slowest and fastest performances. The average computation costs
of \alg~\ref{alg:model-1} and \alg~\ref{alg:model-2}, which embody the
\sda, turn out to be very similar -- within each other's standard
deviations -- and basically identical in comparison with the cost of the
scheme based on the standard test~\eqref{eq:test_above_threshold}. Both
are slower than the hybrid scheme with the standard test: from around
$33\,\%$ slower in the case of high-activity regime with frequent incoming
spikes ($\mu = 18\umV$, $\yj=0.1\umV$), to around $8\,\%$ slower in the case
of low activity regime and infrequent incoming spikes ($\mu = 10\umV$,
$\yj=5\umV$). These are the extremes shown in
\Ref{fig:performance-comparison}. In most other sets of network parameters
the hybrid scheme with \sda\ was around $20\,\%$--$25\,\%$ slower than the
standard hybrid scheme with threshold-crossing
test~\eqref{eq:test_above_threshold}.

As mentioned in \ref{sec:region_missed_spikes_2D}, subregion $\Sp_2$
contains all states for which the standard threshold-crossing test misses a
spike. The occupation frequencies of this subregion are therefore a direct
measure of the number of spike missed, per neuron, by the standard hybrid
scheme. They are shown as red stars in \Ref{fig:region_frequencies} for the
various sets of network parameters $(\mu, \yvar, \yj)$. The frequency of
missed spikes does not have a simple monotonic dependence on the three
parameters, owing to the interaction of several mechanisms, discussed in
\Ref{sec:occupation_freqs}. An increase in synaptic coupling $\yj$ or input
current $\mu$ generally leads to more frequently missed spikes because the
neuron spikes more often overall. For very high -- suprathreshold -- input
currents, however, the frequency decreases again until no spikes are missed
anymore; this change in trend happens because the checkpoints become more
frequent. Regimes of low $\yj$ are diffusion-like processes, where frequent
arrival of synaptic events does not lead to missed spikes. Regimes of high
$\yj$ are shot-noise processes, where sudden and infrequent arrival of
synaptic events leads to suprathreshold excursions and missed spikes.
Separate simulations show that the hybrid scheme based on the \sda, with
either algorithm, reproduce the analytic solution of the neuron model
within floating point precision.

For biologically realistic synaptic couplings, $\yj < 1\umV$, the standard
hybrid scheme is $33\,\%$ faster than the scheme with the \sda, and misses
less than $1$ spike every $10^6$ test calls, per neuron; for very low
couplings $\yj < 1\umV$ this figure even becomes less than $1$ spike every
$10^8$ test calls. For synaptic couplings $\yj > 1\umV$ the standard hybrid
scheme starts to miss more spikes, reaching even $1$ missed spike every
$500$ test calls for subthreshold average input currents; and it is only
$8\,\%$ faster than the hybrid scheme with the \sda. From these figures a
user can decide to use the hybrid scheme with the standard or the lossless
test, depending on the desired balance of accuracy and speed.

\clearpage
\section{Summary and discussion}
\label{sec:summary_discussion}

We here present a general method to solve the threshold-crossing
detection problem for an integrable, affine or linear neuronal
dynamics.  The method is based on the geometric idea of propagating
the threshold plane backwards in time and to determine whether the
swept volume contains the initial state, rather than propagating the
initial state forward in time to check whether it crosses the
threshold. These two procedures are obviously mathematically
equivalent, but they distribute the computational load in different
ways. The forward-propagation of the state looks for the value of the
crossing time; if no such value exists, it means there is no threshold
crossing. The backward-propagation of the threshold first tests
whether a crossing time exists at all, without yielding its value; the
latter is calculated afterwards.

The different distribution of computational load in the two procedures can
be explained geometrically and algebraically. The first procedure
geometrically checks for the intersection of a curve (the state trajectory)
with a hypersurface (the threshold). Algebraically, this corresponds to
finding the roots of a system of equations, often transcendental. The
second procedure geometrically checks for the \enquote{intersection} of a
point (the state) with a hypervolume (the threshold trajectory), \ie\ the
inclusion of the former in the latter. Algebraically, this corresponds to
testing a set of inequalities. The latter procedure is more efficient for
numerical computations, because it only relies on inequality tests -- in
which the presence of transcendental functions is much less costly than in
an equation that needs to be solved for a particular variable. The
determination of the exact crossing-time, which involves bisection
algorithms and is the costlier part, is done only when the existence of
this value is certain. No root search is unnecessarily performed.

We have calculated the system of inequalities expressing the
threshold-crossing condition, for a generic affine or linear neuronal
dynamics in any dimension. The result is the conjunction of inequalities
\eqref{eq:nospike_final_inequalities}. It consists of two affine linear
inequalities in the state-space variables, voltage and currents, and a
non-linear one. The numerical implementation of this system of inequalities
can be further optimized, on a case-by-case basis. In order to give a
concrete implementation example of the generic inequalities
\eqref{eq:nospike_final_inequalities}, to show their geometrical meaning,
and to give an example of optimization,in the present work we apply
our procedure, step-by-step, to the 2-dimensional case of a leaky
integrate-and-fire neuron with exponentially decaying \psc s
\citep{Rotter99a}. The generic inequalities
\eqref{eq:nospike_final_inequalities} take in this case the concrete form
\eqref{eq:main_inequalities_2D}.

The quantitative data in the present work are obtained by integrating
these inequalities into a combined event-and-time-driven simulation
framework \citep{Morrison06c} for large-scale spiking neuronal network
models as released by \cite{nest2015}. Implementation and comparison
to earlier work show that:
\begin{itemize}
\item the system of inequalities, the non-linear one in particular, can be
  expressed analytically in terms of the state-space variables even when
  the original threshold-crossing condition involves a transcendental
  equation -- and would therefore require bisection algorithms. Compare
  \eqref{eq:main_inequalities_2D} with
  \eqref{eq:threshold_cross_time_example};
\item the computationally expensive non-linear function in the system
  can be conveniently triangularized, speeding up the algorithm even
  further by testing a linear inequality first, ruling
  out the majority of
  initial states;
\item the new method reproduces the analytic solution of the neuron model
  within floating point precision. It detects all threshold crossings, in
  particular those that the approximate
  test~\eqref{eq:test_above_threshold} of \citet{Hanuschkin10_113}, misses in
  some ranges of mean activity, fluctuations, and synaptic-coupling strength
  (\Ref{fig:region_frequencies});
\item at the default spike accuracy of $0.1\ums$ of the reference
  simulator the new method is $8\,\%$--$33\,\%$ slower than the
  fastest available solver with spike loss \citep{Hanuschkin10_113}.
  It is therefore of comparable speed as embedded event driven
  methods (see \citet{Hanuschkin10_113} Fig 5, inset).
\end{itemize}

In practice the method of \citet{Hanuschkin10_113} rarely misses
spikes for biologically realistic synaptic amplitudes. At low
frequencies of afferent synaptic events, however, such missed
threshold-crossings can happen. Our new scheme therefore offers an
alternative for users who need guaranteed spike detection and are
willing to pay a price in terms of slightly longer computation time.

In state-spaces of higher dimensions it is be more difficult to derive
the non-linear inequality of the
system~\eqref{eq:nospike_final_inequalities} in implicit form, but its
set of approximating flat surfaces can still be easily calculated. In
the worst-case scenario of an inequality not expressible analytically,
it is still possible to construct a nested sequence of approximating
flat surfaces to be tested hierarchically. Such construction only
needs to be done once for any given model and detects spikes with any
desired precision. Being linear, such nested inequalities likely are
computationally less expensive than a bisection algorithm.

The problem of detecting some sort of threshold crossing in a system of
coupled first-order linear differential equations appears in many other
applications and phenomena like switching, friction, and saturation
\citep{Hiebert80}. For example, in an air-conditioning unit a thermostat
controls the on-off state based on a certain threshold value of the room
temperature \citep{Shampine94}. The dynamics of this system is similar to
that described in Sections
\ref{sec:example_model}--\ref{sec:terms_block_form}. Another example is the
problem of ejecting a pilot such that collision with the aircraft
stabilizer is avoided.

The present work use concepts from differential geometry, in
particular extrusions \citep{bossavit2003b} and critical points of
maps between manifolds, and shows that these concepts have a
readily understandable geometrical and visual meaning. The notion of
extrusion has recently found applications in numerical and
discretization techniques for partial and integral differential
equations \citep{desbrunetal2005}. The notion of critical points of a
manifold mapping is ubiquitous in science: from the caustics of
propagating seismic fronts, at which the seismic wave changes its
phase \citep[\sect~1.04]{romanowiczetal2007}, to the singularities
between two coordinate charts in general relativity
\citep{misneretal1970_r2003}, which affect the accuracy of global
navigation satellite systems \citep{colletal2012,saezetal2013}.

Indeed, the neuron model analysed in \Ref{sec:implementation_example}
exhibits a similarity with the dynamics of a point mass near a black
hole.  If the simulation timestep $\yh$ is very large the curved
surface separating the states that lead to a spike from those that do
not acts like an event horizon in general relativity: a state evolved
from the \spike\ region can enter the \nospike\ region, but once there
it cannot escape and will always remain a \enquote{\nospike}
state. This is only true for the dynamics~\eqref{eq:linear_evolution},
though, with constant affine term and no resets at threshold. Inputs
from other neurons lead to discontinuous changes in the affine term of
the dynamics, causing a \enquote{transport} of initial states out of
the event horizon, from the \nospike\ to the \spike\ region.
Nevertheless, maybe such similarities are more than mere
coincidences. For example, the trajectory of the threshold surface of
a leaky integrate-and-fire model with $\alpha$-shaped post-synaptic
currents \citep{bernardetal1994} can be implicitly expressed in terms of
the Lambert-$W$ function \citep{corlessetal1996}, as an analysis along the
lines of \Ref{sec:maths_prelim} shows. This function also appears in
the implicit expression of point-mass trajectories in $(1+1)$-dimensional
general relativity~\citep{mannetal1997}.
It is surely worthwhile to bring the nascent field of neuronal
dynamics closer to ideas and techniques from differential geometry and
general relativity.

\clearpage
\appendix
\renewcommand*{\thesection}{\Alph{section}}

\section{Interpolation for threshold-crossing time}
\label{sec:membrane_potential_dynamics}

If we know that the neuron voltage $\yV$ is below threshold at times $t$
and $t+\yh$ but above threshold somewhere in the interval
$\opcl{t, t+\yh}$, then by continuity it must reach a maximum above
threshold at a time $\ytm\in \opcl{t, t+\yh}$. This time can be obtained
solving the equation $\Dot{\yV}(\ytm) = 0$, with $\yV(t)$ given by
\eqref{eq:dynamics_2D_model}. The solution \citep{Hanuschkin10_113} is
\begin{equation}
\ytm = -\frac{\tm \ts} {\tm - \ts}\ln\biggl[\frac{\tm}{\ts} - \frac{\tm - \ts}{\tm}  \biggl(\frac{\Ix}{\Is} - \frac{\Cm \yV}{\tm \Is} \biggr)\biggr],
\end{equation}
from which the potential $\yV(\ytm) > \yvt$ can also be easily calculated.

The time $t_{\theta} \in \opcl{t, \ytm}$ at which the first threshold
crossing occurs, $\yV(t_{\theta})=\yvt$, must lie between $t$ and $\ytm$,
with $\yV(t) < \yV(t_{\theta}) < \yV(\ytm)$, and can thus be interpolated
using a bisection algorithm~\citep{pressetal1988_r2007}.

\section{Network setup}\label{sec:noisy_network_setup}

The dynamics of the membrane potential $V$ and synaptic current $\Is$ can
be described, if the input is treated stochastically and for weak synaptic
couplings, by a diffusion process with equations~\citep{fourcaudetal2002}
\begin{equation}
\begin{split} \label{eq:diff-eqn}
\tm \frac{dV}{dt} &= -V(t)+(RI)(t),\\
\ts \frac{d(RI)}{dt} &= -(RI)(t) + \mu + \sigma \sqrt{\tm} \xi(t).\\
\end{split}
\end{equation}
with $\xi$ a zero-mean Gaussian process. The parameters $\mu$ and $\yvar$
characterize the stochastic input and are the mean and variance of the
total incoming synaptic current. They are related to the synaptic couplings
$\set{\yj_i}$ and firing rates $\set{r_i}$ of the input neurons via
$\mu = \tm \sum_i J_i r_i$ and $\yvar = \tm \sum_i {\yj_i}^2 r_i$. If we
have one excitatory and one inhibitory population of input
neurons, mimicked by two Poisson generators with rates $r_{\text{E}}$ and
$r_{\text{I}}$ coupled to the neuron with strengths $\yj_{\text{E}}$ and
$\yj_{\text{I}}$ (where $J = \frac{\tau}{C} w$ and $w$ is the synaptic weight
of the current), and by an input current $\Ix$, then
\begin{equation}\label{eq:mu_sigma_J_r_dependencies}
\begin{split} 
    \mu &= \frac{\Ix \tm}{\Cm} + \tau (J_I r_I + J_E r_E) \\
    \yvar &= \tm({\yj_{\text{I}}}^2 r_{\text{I}}
    + {\yj_{\text{E}}}^2 r_{\text{E}}),\\
    \frac{1}{r}& = \tauref + \tm \sqrt{\pi} \int_{\frac{\Vres-
        \mu}{\sigma} +
      \frac{\abs{\zeta(1/2)}}{\sqrt{2}}\sqrt{\frac{\ts}{\tm}}}^{\frac{\yvt-
        \mu}{\sigma} +
      \frac{\abs{\zeta(1/2)}}{\sqrt{2}}\sqrt{\frac{\ts}{\tm}}} \e^{y^2}\,
    [1+\erf(y)]\,\di y,
\end{split}
\end{equation}
where $r$ is the output firing rate of the neuron, approximated to linear
order in $\sqrt{\ts/\tm}$, $\tauref$ is the refractory time, $\Vres$ the
reset voltage, and $\zeta$ the zeta
function~\citep{abramowitzetal1964_r1972}.

\section*{Conflict of Interest Statement}

The authors declare that the research was conducted in the absence of any commercial or financial relationships that could be construed as a potential conflict of interest.

\section*{Author Contributions}
All authors have made significant, direct, and intellectual contribution to the work.

\section*{Funding}
  This work was supported by DFG Grant: GSC11; Helmholtz association: VH-NG-1028 and SMHB; EU Grant 604102 (HBP); Juelich Aachen Research Alliance (JARA).

\section*{Acknowledgments}
PGLPM thanks the Forschungszentrum librarians for their always prompt and
kind help, Mari \& Miri for continuous encouragement and affection, Buster
Keaton for filling life with awe and inspiration, and the developers and
maintainers of \LaTeX, Emacs, AUC\TeX, MiK\TeX, arXiv, biorXiv, PhilSci,
Python, Inkscape, Sci-Hub for making a free and unfiltered scientific
exchange possible.

\bibliographystyle{frontiersinSCNS_ENG_HUMS} 

\newcommand{\citein}[2][]{\textnormal{\citep[#1]{#2}}}
\newcommand{\citebi}[2][]{\citep[#1]{#2}}
\newcommand{\subtitleproc}[1]{}

\bibliography{portamanabib,
abbreviations,
brain,
math,computer,additional_refs}

\end{document}